# Morphomics via Next-generation Electron Microscopy

Raku Son[1,2)], Kenji Yamazawa[3)], Akiko Oguchi[1,2)], Mitsuo Suga[4)], Masaru Tamura[5)], Yasuhiro Murakawa[1,6,7)], Satoshi Kume[8,9,10)]*

**Affiliations**

1. RIKEN-IFOM Joint Laboratory for Cancer Genomics, RIKEN Center for Integrative Medical Sciences, Yokohama, Japan
2. Department of Nephrology, Graduate School of Medicine, Kyoto University, Kyoto, Japan
3. Advanced Manufacturing Support Team, RIKEN Center for Advanced Photonics, Wako, Japan
4. Multi-Modal Microstructure Analysis Unit, RIKEN-JEOL Collaboration Center, Kobe, Japan
5. Technology and Development Team for Mouse Phenotype Analysis, RIKEN BioResource Research Center, Tsukuba, Japan
6. Institute for the Advanced Study of Human Biology (ASHBi), Kyoto University, Kyoto, Japan
7. IFOM–the FIRC Institute of Molecular Oncology, Milan, Italy
8. Laboratory for Pathophysiological and Health Science, RIKEN Center for Biosystems Dynamics Research, Kobe, Japan
9. Center for Health Science Innovation, Osaka City University, Osaka, Japan
10. Osaka Electro-Communication University, Neyagawa, Japan

***Corresponding author**
Address: 2-2-3 Minatojima-minamimachi, Chuo-ku, Kobe, Hyogo 650-0047
Phone: +81-78-306-0111
Fax: +81-78-306-0101
E-mail: satoshi.kume@a.riken.jp

## Abstract

The living body is composed of innumerable fine and complex structures and although these structures have been studied in the past, a vast amount of information pertaining to them still remains unknown. When attempting to observe these ultra-structures, the use of electron microscopy (EM) has become indispensable. However, conventional EM settings are limited to a narrow tissue area that can bias observations. Recently, new trends in EM research have emerged that provide coverage of far broader, nano-scale fields of view for two-dimensional wide areas and three-dimensional large volumes. Together with cutting-edge bioimage informatics conducted via deep learning, such techniques have accelerated the quantification of complex morphological images. Moreover, these advances have led to the comprehensive acquisition and quantification of cellular morphology, which is now treated as a new omics science termed 'morphomics'. Moreover, by incorporating these new methodologies, the field of traditional pathology is expected to advance, potentially with the identification of previously unknown structures, quantification of rare events, reclassification of diseases and automatic diagnosis of diseases. In this review, we discuss these technological and analytical advances, which have arisen from the need to analyse nano-scale bioimages, in detail, as well as focusing on state-of-art image analysis involving deep learning.

## 1. Introduction

It is said that 'a picture is worth a thousand words'; in line with this sentiment, scientists have been developing tools and techniques to visualise biological specimens for around 400 years. Since Robert Hooke first published 'Micrographia' with beautiful illustrations of cells and living organisms in 1665 [1], the use of light microscopy has led to many important discoveries, not only of various microorganisms, including Mycobacterium tuberculosis [2] and Treponema pallidum [3], but also of cellular components, such as red blood cells [4], capillary vessels [5, 6], brain neurons [7–9] and intracellular structures including nuclei [10].

In 1932, electron microscopy (EM) invented by Von M. Knoll and Ernst Ruska [11, 12] advanced imaging in the field of biology, with one example being the microscopic observation of Escherichia coli at a magnification of 10,000 times [13]. In the 1940s, EM enabled the discovery of virus and phage particles [14, 15], which stimulated the later development of virology. Initially, the application of EM in biological research was difficult because of the lack of histological techniques [16]; however, with the development of fixing and staining methods using aldehydes and heavy metals [17–30], EM was applied more broadly to histology [31]. The use of EM has also revealed a variety of cellular functions such as autophagy [32], slit structures in kidney glomeruli [33] and undifferentiated cell states in induced pluripotent stem cells [34, 35]. Thus, the technological developments in EM revealed a new world of intracellular nano-metre-scale histology that brought with it major biological insights.

In the last decade, two important trends in EM research have primarily emerged: (1) coverage of a two-dimensional (2D) wide-range field for simultaneous capture of many cells and/or whole tissues at high resolution [36, 37]; (2) three-dimensional (3D) resolution, which provides a volumetric viewpoint and reveals the stereoscopic morphology of whole cells and the intercellular connections in tissues [37–39]. These developments potentially facilitate the handling of large bioimaging datasets and/or the collection of comprehensive morphological data from biological specimens [40–42]. Thus, EM has gained attention as a potential new omics modality. In this review, we discuss the application of 'big data' analysis to nano-scale bioimages and we highlight the use of deep learning for state-of-art image analysis.

## 2. Implications of EM Observations in Biology

### 2.1. EM Observations of Biological Microstructure

EM produces outstanding images of membranous cellular structures that maintain cellular morphology and contribute to intracellular/extracellular functions including intracellular transport, phagocytosis and migration [43, 44]. The cell forms a thin and flexible lipid bilayer boundary that holds both hydrophilic and lipophilic solutes, such as DNA/RNA, proteins, glycogen, lipid droplets and minerals, inside the cell [45]. Lipid membranes are also located in cellular organelles such as the nucleus, Golgi apparatus, mitochondria and endoplasmic reticulum [45, 46]. EM can capture the membrane structures, and the cytologic images show the localisation and distribution of cellular components and/or their cellular dynamics.

In practice, the structure and localisation of cellular components and organelles can be captured at the nano-scale using chemically fixed biosamples and a resin-embedded ultra-thin-sectioning EM method [47, 48] as follows. To preserve the characteristic microstructures of cell membranes and cellular solutes, specimens must undergo double fixation using a glutaraldehyde agent and osmium tetroxide [49, 50]. Osmium staining is used to selectively visualise intracellular structures as the chemical covalently binds to osmiophilic materials, such as unsaturated fatty acids and biomolecules with unsaturated bonds [29, 51, 52], thereby conferring some electron density to the osmiophilic substrates. Additional staining using heavy metals, such as lead [19, 25], acetic uranyl [23, 53], gadolinium [54] and neodymium [55], is also performed to improve the contrast of intracellular components. For the preparation of bulk samples larger than a millimetre in size, an en-bloc staining method with improved penetration of osmium acid [52, 56–59] has been developed in addition to a prolonged staining technique used for large heterogeneous sample preparation, including the sequential osmification and treatment of samples with potassium ferrocyanide [60]. Through metal staining, cell morphology can be observed with a clear membrane contrast and the retention of cellular components.

### 2.2. Ultramicrotome and Resin-embedding Ultra-thin Sectioning Methods for EM

Resin-embedding before sectioning [61] is important to EM. Epoxy resin is widely used for the resin-embedding of dehydrated samples [62–64]. It can permeate into the microstructures of cells and/or tissues [65], and then the resin-infiltrated sample can be polymerised (at ~60–70°C) into a firm plastic that is suitable for ultra-thin sectioning.

EM observations of sections were first developed in 1939 [66]; however, the cutting method was difficult to demonstrate because microtome equipment, knives and embedding agents did not exist at that time. In 1948, Peace and Baker used a modified microtome to produce ~500-nm-thick sections for optical microscopy [67]. Later, Sjostrand developed a microtome with a thermal expansion feed system; it used a heater in the cutting feed mechanism and the expansion of the spindle via heating resulted in 20-nm-thick sections [68]. In addition, Porter and Blum developed a mechanical feed microtome with a minimum feed resolution of ~25 nm [69]; the thin sections they produced were the precursors to those produced using current ultramicrotomes.

Fernández-Moránab subsequently proposed a method for producing ultra-thin resin sections using a diamond knife [70–72]: trimmed resin-embedded samples were mounted on an ultramicrotome and sectioned at room temperature, usually to <100 nm, using a diamond knife with a wedge angle of 35–45° [73, 74]. Currently, commercially available diamond knives for ultramicrotomes are around 2–4 mm in width. It is challenging to grind diamond knives wider than 4 mm using a rake angle ≥45° because the cutting edge must be configured with an acute angle. In addition, the wider blade can easily spill during cutting, leading to traces and phenomena on the cutting surface such as knife marks, striations, chatter, vibration and compression [75], which become artefacts during EM observations.

### 2.3. Cutting Theory for Section Preparation

To maximise efficiency and success during ultramicrotome sectioning, it is necessary to accumulate cutting data as well as considering basic theory of cutting. For example, complex correlations among parameters, such as section thickness, knife angle, clearance angle, rake angle, sectioning speed and Vickers hardness, during cutting determine the quality of ultra-thin sections [76]. To improve machinability, it is necessary to reduce sectioning forces in the XYZ axis direction; this requires quantification of the resistance force during shear processing of the sample, optimisation of cutting conditions and clarification of the hardness of the sample [76].

### 2.4. Automatic Collection of Serial Sections

With the advent of the 3D EM methods, the preparation of serial sections from specimen blocks has become possible. Generally, serial sections are prepared using a conventional ultramicrotome by floating the sections on the water surface using a boat attached to a diamond knife [77]. When several sections have accumulated, they are manually scooped out of the water using a glass or silicon substrate. However, it is currently difficult for even expert technicians to manually prepare and collect several hundred serial sections without missing pieces, and the process has low reproducibility. Furthermore, collecting a large number of sections is time consuming [78] and the position of the collected sections varies. Consequently, new serial-sectioning methods have been proposed. One such method involves the use of an automated tape-collecting ultramicrotome (ATUM), a device that automatically collects serial tissue sections using an ultramicrotome and magnetic polyimide tape [79, 80]. The ATUM devices are mainly used in connectome research, which aims to elucidate the network of neurons in the entire brain [81, 82], but they may be used in other research fields. Recently, automated ultramicrotome techniques and diamond knives particularly for use in continuous section preparation have been developed [83]; thus, several hundred to one thousand serial sections can be prepared and collected. Further details are outlined in Section 5.

Current ultramicrotomes are limited in terms the width of sections that can be cut (1–2 mm) and the range of movement of the z-axis of the device. Challenges also exist with high-throughput section preparation and multi-specimen processing. To resolve such problems, the Yagishita Giken Co., Ltd. (Wako, Japan) and RIKEN research groups are developing an ultra-thin section preparation system that can prepare and collect large-area ultra-thin sections by combining machine elements and machining tools for precision cutting. For example, a microtome equipped with a new cutting technique is being developed; in this ultramicrotome system, the aim is to control the cutting feed axis and infeed axis with high repeatability and positioning performance. The recovery system for ultra-thin sections should be able to numerically control the running speed to match the stable tape running speed and the intermittent generation speed of the sections. A synthetic diamond knife has also been developed with a blade width of 10 mm, a blade angle of 50° and stability in terms of material and quality. To obtain stable ultra-thin sections, it is necessary to optimise the processing conditions by observing the cutting surface properties of the specimen and measuring the cutting resistance force. In time, additional researchers from various fields are expected to contribute to morphology research and accelerate the development of technology in this field.

## 3. Transmission EM for Wide-area Imaging in Biology

### 3.1. Conventional Transmission EM and Its Challenges

In biological and medical fields, EM observations typically involve bioimaging of stained thin sections of plastic-embedded samples using transmission electron microscopy (TEM) [84, 85]. In TEM, the device accelerates an electron beam with an extremely short wavelength and irradiates the thin section. Through detection of the transmitted and forward-scattered electrons through the thin section [86], a 2D projected magnified image of the specimen can be obtained at a sub-nano-scale resolution (Fig. 1a). Specimens such as bulk tissues must be sufficiently thin to allow electrons to pass through. The ultra-thin sections are typically ≤100-nm-thin because thicker sections cause inelastic electron scattering [87]. Such sections are placed on a metal mesh grid for observation [88].

Some limitations exist when attempting to prepare ultra-thin slices suitable for TEM observation. First, the metal mesh itself interferes with observations of overlapping tissues [88, 89]. Second, the brittle slices are prone to breakage, which restricts imaging time. Third, since large specimens do not fit in a single field of vision (FOV) of the microscope [85, 90], a controlled system is required to automate imaging process [91] and a handling system is needed for large-sized digital images [42]. These limitations have restricted the use of conventional TEM when observing narrow areas or a relatively small set of cells [84]. Consequently, the current application of EM in clinical diagnosis is limited to assisting diagnosis of renal diseases, undifferentiated tumours, metabolic diseases that mainly affect the muscles or nerves and diseases with unknown aetiologies [85, 92].

Some progress has been made in overcoming the limitations associated with TEM observations. First, a large-sized window and tough supporting film with uniform thickness have been developed to assist with observing wide-range areas [37, 93]. One such supporting film, the LUXfilm® support film, is a highly transmissive and robust film that is better suited for automatic TEM workflows; however, it produces substantial noise without any noise reduction [94]. Konyuba et al. proposed a large-scale silicon nitride window chip deposited using low-pressure chemical vapour deposition as a new support grid for wide-area TEM imaging [95]. This chip is mesh-free, which allows wide-area support for the specimen without creating imaging interference.

A large number of digital TEM images can be captured using an auto-acquisition system with the device [96, 97]. When the physical movement on the microscope stage is not sufficiently precise to obtain the required imaging resolution [98], computational registration and stitching techniques of digital images can be used; these reconstruct single-captured wide-area images from individual tile images [90, 99–107]. These tiled images are also known as montage or mosaic images. Toyooka et al. reported the use of wide-area TEM imaging with a tiled scan of a whole plant cell; this technique successfully produced 3,000–5,000 digital images with the desired range of observation and comprehensive detection of plant organelles [91, 108, 109]. Bock et al. reported electron micrographs of an entire 120,000 × 80,000 pixel thin section of the mouse visual cortex using controlled automated x–y stage motion and image acquisition [37]. Faas et al. performed large-scale EM analysis known as virtual nanoscopy, a methodology for ultra-structurally mapping regions of cells and tissues as large as 1 mm2 at a nano-metre resolution [36]. Lamers et al. imaged human severe acute respiratory syndrome coronavirus 2 (SARS-CoV-2)-infected intestinal organoids autonomously using virtual nanoscopy slides and TEM tomography [110]. As shown in Fig. 2a, through wide-range TEM imaging, including TEM [JEM-1400/Matataki Flash Camera (2,048 × 2,048 pixels), Jeol Ltd., Japan] with a silicon nitride window chip and an automated montage system on Limitless Panorama software (Jeol Ltd., Japan), it was possible to obtain a view of mouse glomeruli that consists of 8,500 tiled images. Using this technique not only preserves the conventional resolution required to capture the basement membrane of the glomeruli and podocyte foot effacement but also enables imaging of multiple glomeruli within the same captured view.

### 3.2. Scanning-type TEM for Large-scale Microstructural Imaging

The nanotomy project (http://www.nanotomy.org/) and its related works provide systematic virtual nanoscopy studies mainly using scanning-type TEM in which the electron probe is scanned across the sample and the transmitted signals are detected point-by-point to form an image [111]. Accordingly, large-scale morphological views of various biosamples, including human pancreas tissue with type 1 diabetes [112], human autoimmune blistering disease pemphigus [113], human skeletal muscle biopsies with histological minimal myositis and capillary pathology [114], SARS-CoV-2 in human tissues [115], human haematopoietic stem/progenitor cells [116], zebrafish brain tissue [117], a free-living marine flatworm [118], rat islets of Langerhans [88, 119–121], mouse salivary gland organoids [122], mouse skeletal muscle [123], mouse kidneys and glomeruli [124, 125] and hamster liver sinusoids [126], have been demonstrated. Recently, Dittmayer et al. developed a methodology for preparing large-scale digitised samples designed to acquire entire sections free from obscuring flaws using scanning EM in the transmission mode; this technique will substantially improve information and throughput gain when analysing experimental and/or clinical samples, including diagnostic muscle, nerve, and kidney samples [127].

Large-scale EM with energy-dispersive X-ray (EDX) analysis enables the acquisition of elemental composition patterns from the surface of samples as well as the visualisation of traditional grey-scale EM images for composition-based interpretation [128, 129]. EDX analysis of the rat pancreas has been used to distinguish, for example, cytoplasmic mitochondria and granules via elemental fingerprinting [120]; thus, analysis of disease tissues may also be possible using composition-based EM. In the objective identification of human pancreas cell types with type 1 diabetes, element maps of granule content produced using EDX analysis have provided data on the elemental variation of granule content within each of the aforementioned cell types [112]. Thus, EDX analysis enables unbiased fingerprinting of cell types and the functionalities of each cell can be inferred from elemental fingerprinting.

## 4. Large-scale Bioimaging Using Scanning EM

### 4.1. Application of Scanning EM in Biology and Recent Progress

Scanning EM (SEM) involves the use of a different type of electron microscope to that used for TEM (Fig. 1b). According to reports, SEM was developed in 1965 [130, 131], about 30 years after TEM. In SEM, the incident electron probe scans across the surface of a specimen in a raster fashion [132], and the interaction between the relatively heavy elements containing the sample and the impacted electrons produces various types of emissions including secondary electrons, backscattered electrons and characteristic X-rays [132]. By detecting such emission types, SEM creates images that reflect the topological contrast or compositional information of specimens as signal intensities in the digital images [132, 133]. Because typical SEM measurements do not require the transmission of electrons through the sample, the process can be used for surface observation of semi-thin sections [134, 135] and bulk specimens [136] such as the surface of kidney glomeruli [137–142], microbial adhesion [143, 144], cell adhesion [145], virus-infected cell surface [146] and blood cells [147–151]. However, traditional SEM images lack the characterisation of internal ultra-structures due to the relatively lower signal under high-magnification imaging conditions [134, 136, 138, 140, 150]. Thus, the application of SEM in diagnostic pathology is limited [152].

Fortunately, recent progress in SEM imaging [153] has led to new SEM utilities, e.g. scanning across ultra-thin sections of resin-embedded specimens under conductive support [154–156]. Scanning electron microscopes equipped with a cold-field emitter [153, 157], a Schottky emitter [158, 159] or a thermal-field emitter [160] as an electron source are known as field-emission scanning electron microscopes; they produce high-resolution images because of the smaller spot size from these emitters [132], the negative stage bias potential [161] and the improved sensitivity of the multiple electron detector system [132, 162, 163], even when the ultra-thin tissue sections are <100 nm. The backscattered electron detection by SEM when using resin-embedded ultra-thin sections provides a reverse contrast of the view that is conventionally possible when using TEM [164, 165]. Although the contrast of the backscattered images is reversed, the quality of the images is sufficient to enable general morphological analysis from TEM observations [85, 166–173]. Furthermore, SEM observations provide high-resolution histological images that are independent of section thickness through regulation of acceleration voltage. Observation of tissue sections through a combination of SEM and resin-embedded sectioning is also known as the section SEM method. In addition, other SEM methods, such as helium ion scanning microscopy [174–177], transmission-mode SEM [178, 179], atmospheric SEM [180] and environmental SEM [181, 182], have been developed.

Recently, the popularity of SEM has lowered the accessibility of the usage and potentially lowered the operating costs [183]. For surface observation with the SEM, various shapes of sample stands can be used provided that they are not made of electrically charged materials. A silicon wafer is a typical base used for biological SEM specimen observation [184]; indeed, huge specimen bases, e.g. 10-cm-diameter wafers, are available. Silicon wafers also adhere well to ultra-thin sections. Sections scooped up on the wafer can be stably stored even when the section is large [50]. In addition, glass slides that are inherently prone to static electricity can be used under conductive treatment by applying a metal coating to the slides before SEM observations are performed [185, 186]. Because SEM has typically been used for surface observation of bulk samples, the sample storage space is designed to have a large XYZ dynamic range. Such advantages of the conventional technique can be fully exploited with the improvement of SEM devices. In other words, the stable fixation of the sample on a sample board has made it possible to observe samples over a long period. The large XY dynamic range facilitates the introduction of relatively large sections (i.e. several millimetres), multiple sample sets and hundreds of sections into the instrument at the same time. The use of the range in the z-direction has made it possible to include microfabrication methods, such as knife cutting and laser cutting, in the sample chamber of the scanning electron microscope [164, 187, 188].

### 4.2. Wide-area Imaging Using SEM

With the aim of practically producing a fish-eye perspective view, early panoramic imaging with SEM was performed in the 1990s [189]. To convert the mosaic images of SEM into a combined image, montage capturing software and image stitching algorithms were developed, similar to existing software for microscopy images [102, 103, 190–195]. For wide-range SEM imaging, Brantner et al. demonstrated large-area and high-resolution mosaic imaging of a 2.5 × 1.8-mm mouse spinal cord resin section (a biologically relevant scale) using the workflow of Chipscanner's laser interferometer stage, FOV mapping and an image stitching technique [196]. Kataoka et al. indicated that stitching SEM enabled the observation of an entire pulmonary alveolus with influenza virus particles in a resin section [197]. More et al. applied a montage SEM imaging technique to quantify the number, myelination and size of axons in the rat fascicle using a computer-assisted axon identification and analysis method [198]. Maeda et al. reported the results of cell counting with autophagy-like vacuoles in wide tissue fields (~600 images in a total area of 0.25 mm2) of the mouse cortex using an automatic acquisition system for tiling SEM images [50]. Kume et al. reported an imaging database of wide-range montage SEM images and their metadata for various tissues, including those from the kidneys, liver and brain cortex region of rodents and human cultured cells [42]. Figure 2b shows imaging data obtained using wide-area montaged SEM images of a rat liver. We integrated more than 1110 images to reconstruct the rat liver leaflet in this large-area image (1 x 0.6 mm). Strikingly, we were able to observe the whole liver lobule while preserving the spatial resolution in EM. The image information obtained using wide-area EM is substantial, which makes interactive visualisation difficult. To solve this problem, image data is converted to the Deep Zoom Image format, which is a layered format with different resolutions on a pyramid structure; this allows interactive zooming in and out for improved visualisation using web software such as Google street view and OpenSeadragon [36, 42, 100]. The use of wide-area EM imaging avoids arbitrary selection of target regions in experimental or diagnostic specimens, and it enables the efficient and comprehensive observation of biological tissues in a time-efficient manner without susceptible bias.

### 4.3. Multibeam SEM

SEM imaging is sometimes more time consuming than TEM imaging due to raster fashion scanning. Thus, methods for speeding up SEM imaging have been developed as follows: (1) image capture with a higher speed single beam, (2) imaging different sections in parallel on multiple EM devices and (3) parallelised imaging of the same section using multiple scanning beams. As a method of parallelised imaging, Eberle et al. demonstrated a throughput imaging technique with multibeam SEM [199–201]. In this system, 61 electron beams are scanned over the sample with one global scanner and secondary electron signals are acquired for each scan position of each beam [199–201]. The multibeam SEM then produces 61 montaged SEM images simultaneously as a hexagonal FOV. In the resultant images, all membranes of neural tissue were clearly visible and intracellular organelles were distinguishable [199]. One hexagonal FOV is used to image a region of about 100 μm2; however, by performing montage imaging, a region ≥1 mm2 can be imaged [199]. Pereira et al. reported that a surface area of 5.7 mm2 could be imaged in a human femoral neck tissue sample, resulting in 897 hexagonally shaped multibeam FOVs comprising ~55,000 high-resolution image tiles and 75,000 megapixels [100]. Multibeam SEM with 196 electron beams has also been developed; this SEM device was designed to detect transmission electrons and backscattered electrons [202]. Thus, multibeam EM systems contribute to high-speed collection of digital images. The applications of multibeam SEM include a much wider-range 2D imaging in addition to 3D EM analysis and brain connectomics research [81, 82, 203, 204].

## 5. 3D Resolution Bioimaging using EM

### 5.1. Implications of 3D Resolution for Ultra-microstructural Observations

To obtain histological and cellular images of targeted 2D regions, the use of ultra-thin section EM techniques with resin-embedded samples is widely accepted and has led to new biomedical discoveries [31]. Indeed, a cellular image obtained from only one tissue section contains substantial biological and medical information. The steric and complex communication of many cells allows living tissues to exhibit and maintain their function [81, 205, 206]. Occasionally, the appearance of characteristic cells and compositions in diseases serves as a biomarker for disease identification [112, 207]. However, the thickness of ultra-thin sections is 50–200 nm; assuming that the actual size of a cell is ~10 μm, one ultra-thin section can be used to interpret cellular events in around one-fiftieth to one-two hundredth of the total cell volume. In most cases, even within the same cell, the shape of the cell nucleus differs greatly depending on the cutting angle and position of the cross section (Fig. 3a). In other words, when a cell image is observed in a cross section, it is difficult to precisely describe whole-cell morphology. In addition, there is an ongoing debate among researchers as to whether the cellular view obtained from extremely thin sections contains artificial cutting bias such as compression. In such cases, visualising the entire morphology of target cells or tissue regions in 3D resolution is required. To realise 3D-directed resolution in EM techniques, observing multiple tomographic images for each cross section one-by-one is a reasonable method [208, 209]. Thus, it is expected that the generalisation of stereoscopic EM techniques will lead to a deeper understanding that would otherwise not be obtained using conventional 2D EM techniques. Here, we reviewed the morphomics techniques used to obtain volumetric EM images (Fig. 4).

### 5.2. Focused ion beam SEM

Focused ion beam SEM (FIB-SEM) is used to observe the surface of a specimen milled by an ion beam on the sample stage of the scanning electron microscope [210]. By repeatedly and alternatively exposing and imaging the new top surface, serial images are captured, although the cutting surfaces cannot be preserved (Fig. 4a). FIB-SEM offers the best z-axis resolution at 4–5 nm; thus, it is suitable for mesoscale observations such as for the observation of cellular organelles [211, 212].

FIB-SEM was originally applied in the 1990s [213], at which time the area covered by ion beams was far smaller than it is today. As the area of observation is enlarged in FIB-SEM, it is commonly applied to various biological samples. Moreover, the outstanding z-axis resolution of FIB-SEM has seen it applied for observations of intracellular events and organelles [214] including in yeast [215], the rodent brain [164, 212, 216], mouse scleral fibroblasts [217], mouse bone marrow adipocytes [218], mouse periodontal ligaments [219, 220], rat kidney glomerular endothelium [221], rat glomerular podocytes [206, 222], human skin fibroblasts [223], human lung epithelium [224], glandular epithelial cells [210], mammary gland organoids [225] and primary mouse pancreas β cells [226]. Using FIB-SEM, Miyazono et al. demonstrated dramatic mitochondrial structural changes that were triggered by the loss of mitochondrial membrane potential [227]. Notably, Xu et al. enhanced the FIB-SEM system by accelerating image acquisition; the speeded-up system allowed imaging of a Drosophila brain at 106 μm3 [228, 229], which serves as a powerful dataset in brain connectomes. Furthermore, Xu et al. reported volumetric image datasets of whole cellular architecture with the finest possible isotropic resolution (about 4 nm square voxel) using FIB-SEM [230], provided as open access data via OpenOrganelle (https://openorganelle.janelia.org/) [214, 230], which allows the study of comprehensive cell morphology [226, 231].

### 5.3. Serial Block-face SEM

Serial block-face SEM (SBF-SEM) is used to observe an exposed sample surface cut using a built-in diamond knife [232]. Compared with FIB-SEM, SBF-SEM facilitates the handling of a broader area as well as faster sample sectioning. SBF-SEM produces 1,000 3D EM images, but it cannot preserve processed sections (as with FIB-SEM) (Fig. 4b).

The prototype of SBF-SEM was produced in 2004 by Denk et al. [233]. They not only showed the power of the technique to reconstruct 3D tissue nano-structures but also directed visualisation of neural circuit reconstructions in neuroscience research [164, 234]. In brain research, the largest mammalian cerebral cortex dataset yielded a reconstruction ~300-fold larger than that in previous reports, which allowed the analysis of axonal patterns [235]. In kidney analysis, Ichimura et al. revealed novel 3D structures in rat podocytes [206, 236]. Other reports also have shown the feasibility of SBF-SEM in 3D EM studies including studies of Drosophila epithelium [237], zebrafish blood vessels [238, 239], mouse B cells [239, 240], rat livers [205] and mouse and human kidneys [241–244]. SBF-SEM has also been applied to image the whole structure of yeast [245, 246], Trypanosoma [247] and cultured cells [248–251].

### 5.4. Array Tomography

Array tomography (AT) is also used to achieve stereoscopic EM (Fig. 4c). In the AT method, serial ultra-thin sections are prepared from a resin-embedded block using an ultramicrotome and then the same site for each section is observed sequentially using TEM or (primarily) SEM [252]. A continuous tomographic image is then reconstructed to obtain the 3D tomography. Unlike other methods, the AT method is notable for its capacity to preserve thin sections, which could then be re-observed later. The resolution of the z-axis in the AT method is the thickness of the section, which is approximately 50–100 nm. Combining the AT method with SEM potentially allows for wide-area volumetric observations [40]. This technique is also known as the serial-section SEM method [184].

The idea for creating serial sections dates back to the 1950s, soon after the first ultramicrotome became available [77]. Ribbon-like serial sections were transferred to supporting grids for observation. Later, in a 1970s report on 3D construction of the juxtaglomerular apparatus of the rat kidney, 500 serial sections were obtained and TEM observations were successful [208]; however, 3D illustrations were limited. Subsequently, TEM was used for 3D EM, especially until the 2000s. Serial-section TEM (ssTEM) images have been acquired from serial sections of differentiating monocytes [78], neuron connections [253], yeast cells [254] and the endoplasmic reticulum [255]. Notable results of AT and TEM combined include whole-imaging of an adult Drosophila brain using a custom high-throughput ssTEM platform developed by Zheng et al. [256]. This volumetric morphology obtained by ssTEM has contributed to mapping brain-spanning circuits and accelerated research in the field of neuroscience.

In parallel with the development of ssTEM, AT combined with SEM was proposed in 2007 [257]. This combination method has been used to study varicella-zoster virus-infected cells [258], the Golgi apparatus in different cell types [259] and sorted immune cells [260]. In addition, we could successfully generate a 3D volume EM image of a human leukaemia cell and the macula densa in the distal tubules of a mouse kidney glomerulus using AT combined with SEM (Fig 3bc). The SEM-based serial-sectioning method is suitable for relatively large samples because it collects larger-area serial thin sections onto the silicon substrate or glass slide [184, 261]. However, the AT method is generally challenging because it is difficult to manually prepare continuous ultra-thin sections of hundreds to thousands of samples. To improve the AT technique, customised AT methods have been developed such as magnetic collection of ultra-thin sections [203], customised substrate holders [40, 262, 263], a modified AT-boat diamond knife [83], tape collections of sections [79, 81, 264, 265], a carbon nano-tube tape for serial sections [82] and the combination of semi-thin serial sectioning and FIB-SEM [266]. Among these, the tape collection system using ATUM has facilitated automatic collection of tissue serial sections and volumetric SEM [79, 267]. This ATUM-based AT–SEM method was used to clarify the sub-volume of the mouse neocortex from ~2,000 serial sections [81] as well as all myelinated axons of the zebrafish brain from 16,000 serial sections [268]. Morgan et al. imaged 10,000 sections of the mouse visual thalamus, which were produced using ATUM to a thickness of 30 nm, with an imaging volume for the dataset of 0.8 mm × 0.8 mm × 300 μm; 899 synaptic inputs and 623 outputs were mapped in one inhibitory interneuron [269]. Witvliet et al. used the ATUM–SEM system to reconstruct the full brain of eight isogenic Caenorhabditis elegans individuals across postnatal stages in an age-dependent manner [270], which provided insights into the mechanism of connectome development during brain maturation. These obtained datasets are also provided in BossDB (Brain Observatory Storage Service & Database, https://bossdb.org/) [271–273]. Recently, Shapson-Coe et al. applied a combination of ATUM-based AT and multibeam SEM to petabyte-scale large volume imaging of the temporal lobe of the human cerebral cortex, and they computationally rendered the 3D structure of 50,000 cells, hundreds of millions of neurites and 130 million synaptic connections in the volumetric images; their findings suggested the existence of a new subset classification of deep-layer excitatory cell types [274]. Moreover, the use of such large-scale stereoscopic EM techniques to analyse the microstructures of pathological conditions is expected to improve our understanding of disease-specific structures that could not be obtained using conventional EM techniques.

### 5.5. 3D Imaging using High-Speed TEM

To develop high-throughput TEM imaging, Graham et al. used a tape-based, reel-to-reel pipeline that combines automated serial sectioning and a TEM-compatible tape substrate, GridTape [275]. This acquisition platform provides nano-metre-resolution imaging at fast rates via TEM. Based on this pipeline, multiple-scope parallel imaging using a 50-MP camera has enabled image acquisition of a >1-mm3 volume of the mouse neocortex, spanning four different visual areas at synaptic resolution, in less than 6 months; in turn, this has yielded a >2-petabyte dataset from over 26,500 ultra-thin tissue sections [265]. In addition, Phelps et al. applied GridTape-based serial-section TEM imaging to acquire a synapse-resolution dataset containing an adult female Drosophila ventral nerve cord [264]. The complete connectivity maps provided a deeper understanding of how the nervous system controls the locomotor rhythms underlying swimming and crawling [264]. Since TEM offers much faster imaging compared with that of SEM, this research could be applied in areas that require broad observation with precise imaging.

## 6. Correlative EM and Multimodal Imaging

### 6.1. Correlative Light Microscopy and EM

The body can be understood more deeply if tissue functions and macromolecular fingerprinting can be estimated at the nano-level, which is sometimes difficult to achieve using only EM-based morphomics. For example, distinguishing between excitatory and inhibitory neurons cannot be achieved based only on morphology [276, 277]. To resolve this issue, correlative EM, which combines EM and other imaging tools, can be used to better understand molecular functions and other factors. Correlative EM is also useful for screening or targeting specific structures, especially when targeting cellular markers. A well-established example is the combination of light microscopy and EM, known as correlative light and electron microscopy (CLEM) [278]. The idea behind CLEM was first proposed in the 1980s [213]. Samples are initially imaged using a light microscope to detect histological morphologies or fluorescence signals, after which they are subjected to EM imaging with a nano-resolution. Correlation imaging is achieved either by sharing the same FOV for both modalities or by sample transfer in tandem [278].

To screen for structures of interest, Ronchi et al. developed a workflow that combined fluorescent labelling and FIB-SEM, which enabled correlative targeted imaging of animal mammary gland organoids, tracheal terminal cells and ovarian follicular cells [225]. CLEM has also been applied to the mouse brain [252, 279–283] and the ferret brain [284] as well as whole model organisms [83], cultured cells [227] and various tissues [232]. In addition, CLEM could be used for in vivo multicolour imaging, known as Brainbow [285–287].

When targeting specific structures, Trzaskoma et al. applied CLEM to reveal 3D chromatin folding [288]; they combined DNA fluorescent in situ hybridisation with SBF-SEM. Oorschot et al. published a workflow integrating the Tokuyasu technique to preserve the antigenicity of proteins and investigated neural stem and progenitor cell populations [276]. In addition, 3D CLEM combined with the CryoChem technique allows for quality ultra-structural preservation that is broadly applicable to cultured cells and tissue samples [289]. The CLEM method can even be used to target specific proteins under transgenic conditions via the engineered peroxidase gene APEX2 [290–293], which serves as a labelling probe in both EM and light microscopy [294]. This APEX2 system has been successfully implemented to track lysosomes in dendrites [295] and to visualise the localisation of endoplasmic reticulum chaperonin [296], the outer endoplasmic reticulum and mitochondrial membrane [297], membrane proteins [298] and multicolour labelling of peroxidases [299]. Recently, the APEX-Gold method, which has high sensitivity, was used as a genetic tagging in 3D EM [300].

For correlation of live cell imaging, Fermie et al. analysed the dynamics of individual GFP-positive structures in HeLa cells and then correlated these with images from FIB-SEM [301]. Thus, they overcame the limitations of EM, i.e. that EM cannot visualise live cells. Betzig et al. first introduced a combination of super-resolution light microscopy and EM (super-resolution CLEM) to image specific target proteins in the thin sections of lysosomes and mitochondria [302]. Currently, super-resolution CLEM can achieve a resolution of 20–50 nm, although the distortion of the sample becomes a problem at <10-nm resolutions [303]. This technique was also utilised to visualise the Golgi apparatus [304, 305], mitochondria [306, 307] and other organelles [306].

### 6.2. Correlative X-ray Computed Tomography and EM

X-ray computed tomography (CT) has been applied to biological tissues or cells to obtain almost single cell-level morphology data [308]. In practical terms, observing cells that are approximately 10 μm in size requires sub-micro-resolution potential in the CT device. When observing intracellular structures, the use of synchrotron radiation X-ray is necessary. Although, at present, single-cell imaging with a CT device remains a special case, in this section we discuss cellular tissue analysis, including single-cell imaging, using CT devices and further correlative CT-EM.

The following are a few examples of X-ray CT applied in biology research to date. In the early 1980s, the micro-CT technique was developed to achieve 3D observations at micrometre resolutions [309]. This technique can be used to obtain a projection image of a sample by irradiating it with X-rays with wavelengths of ~1 pm to 10 nm. Compared with the 1–2 mm resolution in conventional medical CT scans, micro-CT tomography results in a higher spatial resolution of 1–50 μm (generally approximately 5 μm resolution per voxel) [310, 311]. Because the spatial resolution depends on the focal spot size of the X-ray source [312, 313], relatively small sample pieces (<10 mm in size) can be used for micrometre-level resolution with micro-CT. Moreover, micro-CT imaging provides high contrast results, especially in tissues with high or low X-ray permeability (e.g. the lungs or bones, respectively), without the need for special sample preparation. However, particularly in soft tissues, including the brain and renal cortex, suitable staining techniques are required to increase absorption-based contrast of tissue structures [314–316]. In several studies, micro-CT has been applied to visualise juvenile coral [317], small organisms [318–321], nano-material in lung tissues [322], mammalian brain tissue [310, 323], rodents kidney nephrons [316, 324, 325], mouse liver structures [326], mouse embryos [327–332] and human placenta [333]. In addition, the phase-contrast approach of X-ray CT can be applied generally to unstained specimens such as mouse kidneys [334], the human heart [335], human brain tissue (cerebellum) [336] and plant germination [337]. Overall, this modality has become a promising method used in morphomics.

For 3D non-destructive targeting of a region of interest in a specimen, EM analysis correlated with the X-ray CT modality has been proposed. Several reports of correlative CT and other microscopic techniques have included EM [338–341]. In particular, correlative micro-CT and EM has been applied to clarify neural 3D structures in mouse brain tissues [342]. Silver impregnation staining applied to neurons can also feasibly be used in correlative workflows [343, 344]. Karreman et al. demonstrated the in vivo tracking of single tumour cells using multimodal imaging including X-ray CT and EM [345], which is expected to have broad applications in various biological fields.

Some CT devices even offer sub-micro-resolution [346], i.e. nano-CT, which is sometimes used as a synchrotron radiation-based CT setting and a use of soft X-ray with relatively low penetrating power. Nano-CT has been applied to reconstruct the neural network in Drosophila or the rodent brain [347–349]. In addition, this resolution can achieve cell-level observations [316, 326] that could further facilitate correlative analysis in combination with EM. Interestingly, Kuan et al. demonstrated that X-ray holographic nano-tomography can be used to image large-scale volumes with sub-100-nm resolution in Drosophila melanogaster and mouse nervous tissue, thereby enabling a close reproduction of the EM images [350]. Moreover, multiple scanning technique can comprehensively catalogue mechanosensory neurons and trace individual motor axons from muscles to the central nervous system [350]. Nano-scale X-ray CT can then bridge a key gap that helps move toward EM resolutions. Furthermore, the integration of nano-scale CT and EM has been used to study mitochondrial morphology in relation to drug resistance in human colon carcinoma cells [351]. A parallel-beams CT method can achieve much faster image acquisition and comparable or even better resolution [352]. In addition, synchrotron-based CT has been used for cellular-level analysis of bacteria [353], yeast [354, 355], mammalian cells [355, 356], neuroanatomy [357], renal microvasculature [358] and human bones [359, 360]. Overall, the X-ray CT modality has the potential to be used not only for regional targeting prior to correlative EM analysis but also for morphomics analysis of parenchymal morphology with nano-scale resolution.

## 7. Large Bioimage Datasets and Comprehensive Image Analysis

### 7.1. The Morphome and Morphomics

The morphome or biological morphome refers to the totality of the morphological features in a species [42, 361–364]. The morphome is expressed as the sum of a species' molecular dynamics [362, 363] including its DNA (genome), gene expression (transcriptome) and metabolic (metabolome) information such as lipids and sugars (Fig. 5). It also refers to morphological phenotypes. Most morphological data are imaging data, which at first glance differs from the genomic sequences and numerical data that are mainly used as omics data in molecular biology [364]. Thus, different approaches are required to handle morphome data.

As EM device technologies have advanced [132], the acquisition speed of EM imaging has dramatically accelerated and EM-based imaging techniques have been used to study the complexity of organisms at high-level 2D and 3D resolutions. Indeed, imaging data can be produced at a level comparable with genome data obtained via next-generation sequencing. For example, wide-area imaging produced using single-beam SEM can acquire several tens of gigabytes of data in a single day of imaging [42], whereas 2D/3D imaging produced using multibeam SEM can acquire hundreds of gigabytes or a terabyte scale of image data [201]. Moreover, the latest EM methods, such as high-speed TEM methods [265] and ATUM-based AT and multibeam SEM methods [274], can generate petabyte-scale image data.

These levels of imaging data can be used to systemically measure and quantify large morphological fingerprints and diverse biological phenomena. Further quantitative morphology analysis could be applied to study biological functions. Such comprehensive approaches have led to the treatment of the resultant large imaging datasets as new omics information, which is termed morphomics (Fig. 5). This involves the integration of comprehensive (big) morphology data and bioimaging informatics, which will result in the discovery of unknown features, but there is still a bottleneck of imaging data mining. In the following subsections, we discuss such morphome analysis, outline imaging data operations and highlight image data analysis using deep learning.

### 7.2. Toward the Handling of Massive Bioimage Datasets

Just over 20 years ago, film photographs were still in mainstream use as EM images [365], whereas the EM images of today are high-resolution digital images. In this period of development, advances in infrastructural technologies, such as digital image-archiving, faster network communication, improved computing performance and increased storage disk capacity, have facilitated the acquisition of large-scale digital bioimage data and enabled the practical handling and processing of images [366]. It is now possible to operate with hundreds of gigabytes or terabytes of images even in a laboratory setting. Efforts are being made in the field of bioimaging data operations to handle such huge image datasets for data repositories, data sharing and reuse in a standardised manner [367].

Open-source data and data accessibility are critical to the sharing of bioimaging data [368]; fortunately, worldwide access to data has become possible via the Internet [369, 370]. However, it is necessary to construct a descriptive format and data repository, so-called metadata and an image database, respectively, prior to distributing bioimaging data [371]. The Open Microscopy Environment (OME) consortium is working to produce imaging metadata and public image archives in the medical and life sciences [372, 373]. The OME is an open-source software framework developed to address standards for sharing image data and analysis results [368, 374]. Within its framework, OME Remote Objects, an open-source interoperability toolset for biological imaging data [375], and OME metadata [376–379] have been developed to manage multidimensional and heterogeneous imaging data mainly from light optical microscopy. However, standardised metadata that describes EM experiments, including bioresources, measurement conditions and image formats, has yet to be developed; thus, integrated analysis of imaging data with other metadata has remained difficult [380]. Therefore, we previously proposed the development of microscopy metadata to describe EM experiments and their image datasets based on the data model of OME metadata [369, 380], and we offered a combination of an ontology-based imaging metadatabase and an image viewer, which were distributed in a machine-readable web form [42]. At present, metadata arrangements for bioimaging, including EM, have been discussed internationally toward the reuse of microscopy data [381]. In future work, the application to medical imaging research, such as MRI and PET/CT, for human subjects will also be important [382, 383].

In May 2021, the Global BioImaging (GBI) consortium proposed criteria for globally applicable guidelines related to the tools and resources of open image data in the fields of biological and biomedical imaging [367]. The GBI also founded international non-boundaries to develop common imaging and data standards that promote data sharing and open data, as well as world-class training programmes and repositories of image data analysis tools for use by imaging scientists. Furthermore, the 'Quality Assessment and Reproducibility for Instruments and Images in Light Microscopy' initiative recently proposed the establishment of guidelines for quality assessment and reproducibility related to microscopy instruments and images [384]. These activities are expected not only to improve the overall quality and reproducibility of data across the microscopic bioimaging field but also to enable handling of huge bioimage datasets in a standardised manner. In addition, the barriers to data sharing of EM images should also be reduced.

As a prototype of the first open online repository to link imaging and molecular data, Williams et al. launched the Image Data Repository (IDR; https://idr.openmicroscopy.org/) in 2017 [373]. This platform stores bioimage data from several imaging modalities, including multidimensional microscopy and digital pathology [373], while integrating imaging studies and phenotypic information. It can be searched according to the metadata or image attributes and consists of two major categories: Cell-IDR and Tissue-IDR. Currently, the IDR platform distributes some EM datasets including cultured cell chromatin [385], intestinal organoids [110], budding yeast [386] and zebrafish embryo sagittal sections [36].

As a public archive of EM images, the Electron Microscopy Data Bank [387, 388] was launched in 2002 to provide a public repository of mainly electron cryo-microscopy volume maps and tomograms including macromolecular structures, such as proteins and their ligand complexes, and subcellular structures. In 2016, the Electron Microscopy Public Image Archive (EMPIAR; https://www.ebi.ac.uk/empiar/) [389] was published. This archive became a public resource for raw images underpinning 3D cryo-EM maps and tomograms. Most EMPIAR datasets are particle images and 3D tomograms of macromolecules obtained using cryo-EM, but they currently contain several 3D EM datasets of epoxy-embed tissue and cell samples obtained using SBF-SEM or FIB-SEM. For example, a 3D imaging dataset of the HeLa cell line obtained using SBF-SEM (EMPIAR-10094) [251, 390] consists of 518 cross-sectional images with a size of 8,192 × 8,192 pixels; the dataset is nearly 130 gigabytes in size, indicating that the EMPIAR covers a wide range of biological samples. Notably, all data archived in the EMPIAR is under CC0 licence and can be re-used freely without any conditions or restrictions.

To date, the development of these public bioimage resources is at an early stage and further accumulation of imaging data and development of integrated platforms is highly desirable. Representative list of current public EM datasets is summarised in Supplementary Table 1. In the past, providing large-scale open-source genomic sequence data helped advance our understanding of genomics. Accordingly, increasing the availability of imaging datasets will further stimulate imaging research and the development of novel imaging technology.

### 7.3. Bioimage Analysis Using Deep Learning

Little progress has been made in the methods used to analyse EM bioimages over time; conventionally, every image was examined manually [391]. EM bioimages possess a low signal-to-noise ratio, black-and-white contrast and a variety of morphological features. When using classical image analysis methods, it has been difficult to recognise and decode EM bioimages. For example, in semantic segmentation, which is used to extract a particular region in an image, the use of classical deductive methods [392–394] has failed to identify a mathematical solution for the morphological features of a particular region among various other morphologies. In most EM research cases, comprehensive quantification, including automatic segmentation, could not be achieved even after acquiring large-scale image datasets. In an attempt to resolve these issues, current best practice is to apply cutting-edge informatics or artificial intelligence (AI) approaches [395, 396] to quantify the microstructures in EM bioimages.

Image analysis via AI techniques, such as machine learning (ML) and deep learning (DL), has received substantial attention in the fields of biomedicine [397–399] and imaging research [400, 401]. Inductive analysis using supervised data has been used to identify characteristic changes in morphology [402]. Initially, such AI techniques were applied in neuroanatomy research. Kaynig et al. also demonstrated fully automatic 3D segmentation of thin, elongated, cell membrane structures of dendrites for 30 sections in TEM images by finding features from the images and constructing a features classifier using a random forest, a ML method that uses ensemble learning [403]; this tool was supplied as a plug-in of ImageJ/Fiji [404, 405]. Turaga et al. presented a ML algorithm for training a classifier to produce affinity graphs, representing the x-, y- and z-direction information, which can be used to segment the EM images of neural tissue [406]. Subsequently, they reported an affinity graph computation that used a four-layered convolutional neural network (CNN) trained with the supervised dataset; this resulted in 3D reconstructions of neurites with ~90% segmentation accuracy in a 3D EM dataset of rabbit retina tissue [407]. These were the original applications of ML and DL in connectomics studies. Subsequently, AI technologies have become indispensable for bioimage analysis. For instance, DL models, which use an exquisite combination of multi-layered CNN for learning [408, 409], are the current state-of-the-art technology in image recognition [401, 410]; they can extract morphological features, such as the cell body and nucleus, from cellular images via complex networks [411, 412].

For quantification of bioimaging data by DL, the U-Net model was proposed by Ronneberger et al. [413, 414]; the model has encoder and decoder parts with multi-layered CNNs and contracting paths between the encoder and decoder. U-Net targets segmentation tasks in a small number of image datasets with large feature types that are unique to the bioimage dataset. The U-Net model substantially improves performance in segmentation tasks such as cell division tracking and neuronal cell membrane segmentation [414]. Many derivative models, such as FusionNet [415], enhanced U-Net [416], U-Net-Id [417], SCAU-Net [418] and Dual ResUNet [419], have subsequently been proposed that reportedly improve performance compared with that of the original U-Net model. In addition, 3D applicable models have been developed as an extension to 3D volume data [420–422]. Therefore, the U-Net method was a leading technique that has served as a foundation for biological image analysis. For further details, refer to the review by Siddique et al. [423], which comprehensively discusses the U-Net models and derived models.

To use ML and DL techniques universally, user interface tools, such as QuPath [424], Microscopy Image Browser [425], NuSeT [426], UNI-EM [427] and DeepMIB [428], have also been developed. Additionally, the CDeep3M tool can perform image segmentation in cloud computing [429]. A generalist DL model for cellular segmentation, so-called Cellpose, was proposed, which can be precisely applied to the segmentation of cells from a wide range of image types [430]. The DL model has also been ported to R, and typical models for segmentation are available in the ANTsX ecosystem [431]. In addition, we have recently begun distributing supervised bioimage datasets in the R array format that can be used in the analytical workflow in the R environment as the BioImageDbs project via the Bioconductor ExperimentHub platform [432]. Wei et al. introduced the MitoEM dataset, a 3D mitochondria instance segmentation dataset with approximately 40,000 instances and 30 μm3 volumes brain cortices [433].

In contrast to these successes in bioimage data analysis, automated segmentation tasks related to EM images remain challenging because the texture and intensity variation are generally similar [434]. A benchmark report comparing published seven models for EM images showed that their performances were still highly variable [434]. In recent years, however, successful research cases of EM image analysis using DL have been reported, including in neuroanatomy research and from other fields. In neuroanatomy research, the 3D segmentation of neurites using the DeepEM3D model achieved close to human-level performance [435]. Lee et al. reported a residual symmetric U-Net architecture that achieved an approximately 2%–3% error rate for an EM image dataset of mouse neurites [81] in the SNEMI3D challenge; thus, the system surpassed the human accuracy value provided at that time [436]. Januszewski et al. used a flood-filling network (FFN) to trace neurons in a dataset from a zebra finch brain obtained using SBF-SEM; they achieved high-precision automated reconstruction of neurons with a mean error-free neurite path length of 1.1 mm [437]. Sherida et al. reported that the addition of local shape descriptors promotes affinity-based segmentation methods to a level that is on par with that of the current state-of-the-art system for neuron segmentation based on FFN [438]. Other researchers have attempted segmentation analysis of cell bodies and cell nuclei [251, 390, 439, 440] as well as organelles [441–444] in EM images. For multiple segmentation tasks with EM images, the transfer of learning from pre-trained models using the CEM500K dataset was effective for the transferability of learned features, indicating that a large amount of training data is important for encoding bioimages [445]. Comprehensive quantification of multiple organelles in whole cells using DL segmentation has been reported for serial cross-sections of the mouse liver [446] and for cultured cells [231], which suggests the future possibilities for cell biology that may arise from intracellular morphomics.

Importantly, the DL technique has been applied to more than just quantitative analysis. A recent report described a new data reduction and compression scheme (ReCoDe) that converts the raw data from EM images into 100 times more minor data [447]. Furthermore, image generation models, such as CycleGAN [448, 449], have been applied to EM images as denoising [94, 450–452] and image super-resolution techniques [453–456]. The super-resolution DL technique has been applied to an active acquisition pipeline of SEM imaging [457]. Interestingly, image transformation techniques can also reproduce tissue-stained images from unstained or other stained images [284, 458–462]. To expand limited datasets, various proposals have been made for image data augmentation in DL [463]. In future research, these conversion techniques may also be widely applicable to EM images.

### 7.4. Deep Learning Applications to Medical and Functional Images

The use of DL in pathology has also advanced remarkably. Using AI to analyse tissue sections is often referred to as computational pathology [397]. Cases of its application have been reported in relation to the pathological diagnosis of various cancer types, e.g. breast cancer [464–466], bladder cancer [419], renal cell carcinoma [467], non-small cell lung cancer [468], skin cancer [469] and gastrointestinal cancer [470], with such techniques potentially increasing the precision of oncology results [471]. For kidney disease assessments, the relationship between renal histology and the prognosis/severity of renal disease has been examined using image recognition and comprehensive segmentation of the constituent tissues of renal samples including human renal biopsies [472–475].

Intriguingly, recent DL results have been associated with biological functions such as the prediction of gene expression patterns [476, 477] and genetic mutations from histological images [478, 479]. Prediction of genetic mutation patterns in lung cancer [468], breast cancer [480] and haematologic cancer [481] has been achieved using morphology image recognition, and this enabled severity classification. Digital imaging studies have been conducted to explore the relationship between histology and gene expression patterns for the prediction of genetic variation associated with tissue morphology [482], the RNA-Seq profiles of tumours [483] and multifactorial site-specific signatures of tissue submitting sites [484] from whole-slide images, as well as diverse molecular phenotypes identified by the expression of immune checkpoint proteins in tumours [485]. Future developments may lead to the prediction of gene expression at the nano-scale level.

To ensure that DL becomes more widely available, two weaknesses must be addressed. First, the manual annotations required for supervised learning are time consuming and costly. For example, for an EM dataset of 1 million μm3, the cost to manually annotate all images will be 10 million dollars [486]. However, using an unsupervised learning method, which learns from the data without any pre-annotated labels [487], is one means of solving this problem. Although such an approach without annotations is still rare, some studies have included autoencoders for unsupervised learning that have achieved an acceptable performance in histological image recognition [488–490]. Second, the computational process along with the heavy optimisation of many parameters remains a 'black box', which is not interpretable by humans. Currently, the appropriateness of a network model can only be evaluated using the numerical value of its prediction performance. One solution is to use human-interpretable image features to predict outcomes [485]. In addition, post-DL analysis should be directed at determining which biologically meaningful information can be extracted from the quantification of morphological features.

## 8. The Blueprint for Morphomics

The development EM has placed nano-scale imaging data at the centre of morphological analysis. Large 2D and 3D datasets can be acquired with a millimetre-wide range at a nano-metre resolution. However, previous morphological studies involving big data analysis have been limited to analysis of the individual dataset, sometimes of a single dataset of normal tissue without any comparison. Although comparisons of several morphological features have been conducted in simple datasets, it is currently challenging to compare across datasets or among unbiased structural features.

Morphomics currently stands at means to create a 'reference morphology' or reference EM datasets at an early stage in the development of genomics studies. Only if a solid 'reference morphology', or at least a standardised workflow including sample preparation, microscopic settings, data storage, data normalisation and image analysis, is established, will it be possible to conduct larger scale comparative studies that could have major biological implications. The use of bioinformatics methods and imaging databases will accelerate this process. In addition, further multi-omics analysis techniques that can bridge the gap between morphomics and other omics data will be powerful tools. In particular, we expect the development of revolutionary methodologies that combine large-scale EM data analysis techniques with analysis of genomics data such as gene mutation and expression data. By incorporating these new methodologies, the field of pathology is expected to progress rapidly, which might include the identification of previously unknown structures, the quantification of rare events, the reclassification of diseases and the automatic diagnosis of diseases. Furthermore, the amount of data that can be analysed is expected to increase dramatically with the development of automatic AI analysis.

## Conclusion

Wide-area 2D EM imaging along with large-scale 3D EM image acquisition and its reconstruction for biological tissues are currently defined as next-generation electron microscopy techniques; however, these tools are now more commonly being used to produce massive morphomics datasets. To maximise the utilisation of morphomics data, the general use of DL methods and post-DL analysis will be essential for comprehensively quantifying cellular morphology. Overall, these advanced techniques can be expected to deepen our understanding of living tissues and cells.

**Acknowledgments**

This study was partially supported by the RIKEN engineering network, the RIKEN aging project, the JSPS KAKENHI (Grant Numbers 18K19766 and 15K16536), Prof. Osafune memorial scholarship (the Japanese Society of Microscopy) and the Strategic Core Technology Advancement Program (Supporting Industry Program; SAPOIN) funded by the Ministry of Economy, Trade and Industry in Japan. We thank a member of the SAPOIN project in RIKEN for the regular and helpful discussion as well as Ms. Yuka Watahiki (Frontiers of Innovative Research in Science and Technology (FIRST), Konan University, Kobe, Japan) for her technical assistance with cell segmentation. The English language in the manuscript was reviewed by Enago (https://www.enago.jp/).

**Conflict of interest**

The authors declare that they have no conflicts of interest.

**Author contributions**

RS and SK designed the contents of the review article. AO provided the EM dataset. RS, KY, AO, MT, MS, YM and SK wrote the manuscript. RS and SK edited the entire manuscript.

**Figures and Figure legends**

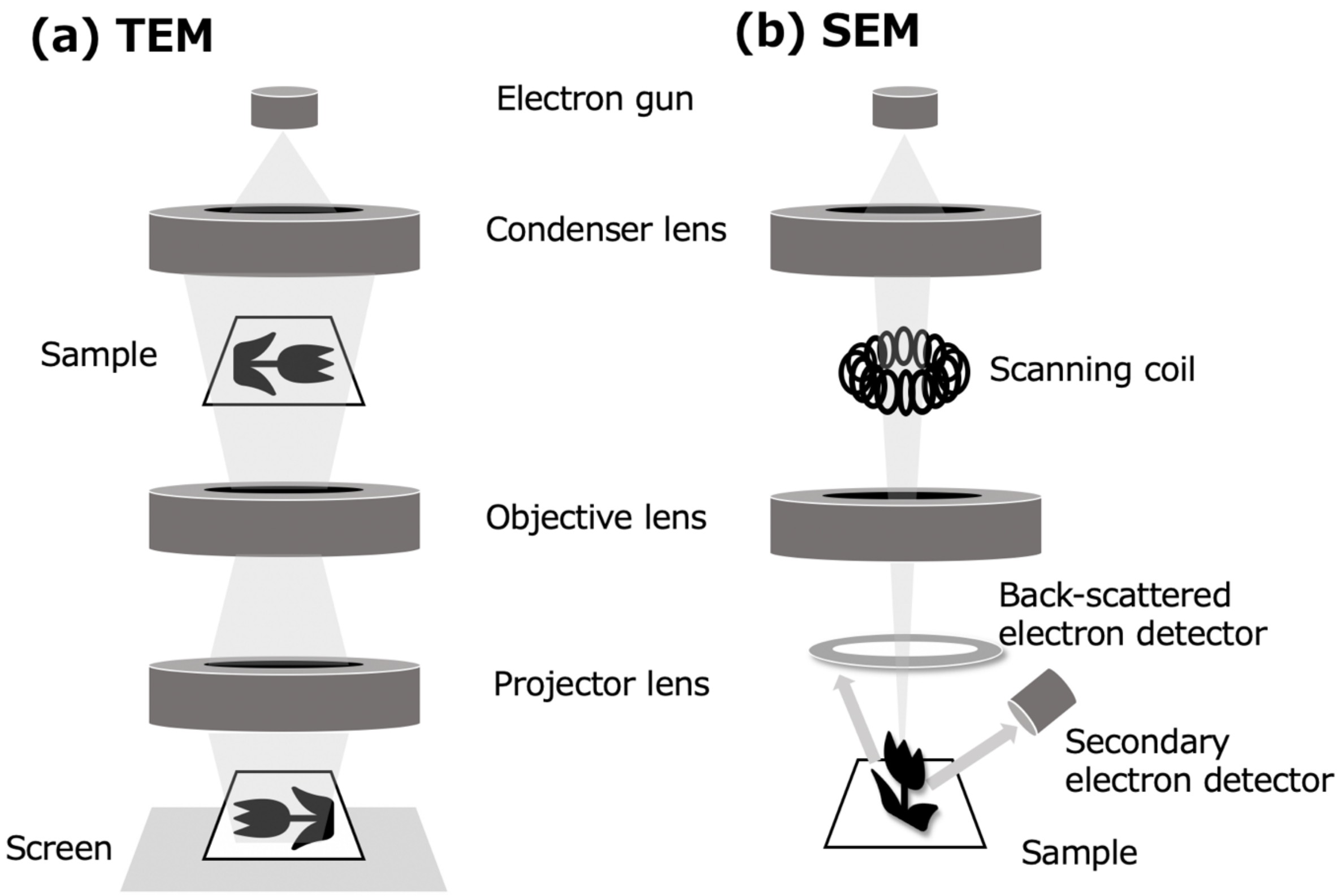

**Fig. 1** Imaging principles in transmission and scanning electron microscopy (TEM and SEM)

(a) TEM captures the forward transmitted electrons through the sample, whereas (b) SEM scans the sample with a narrow beam probe and captures various signals with different energies including the backscattered and secondary electrons.

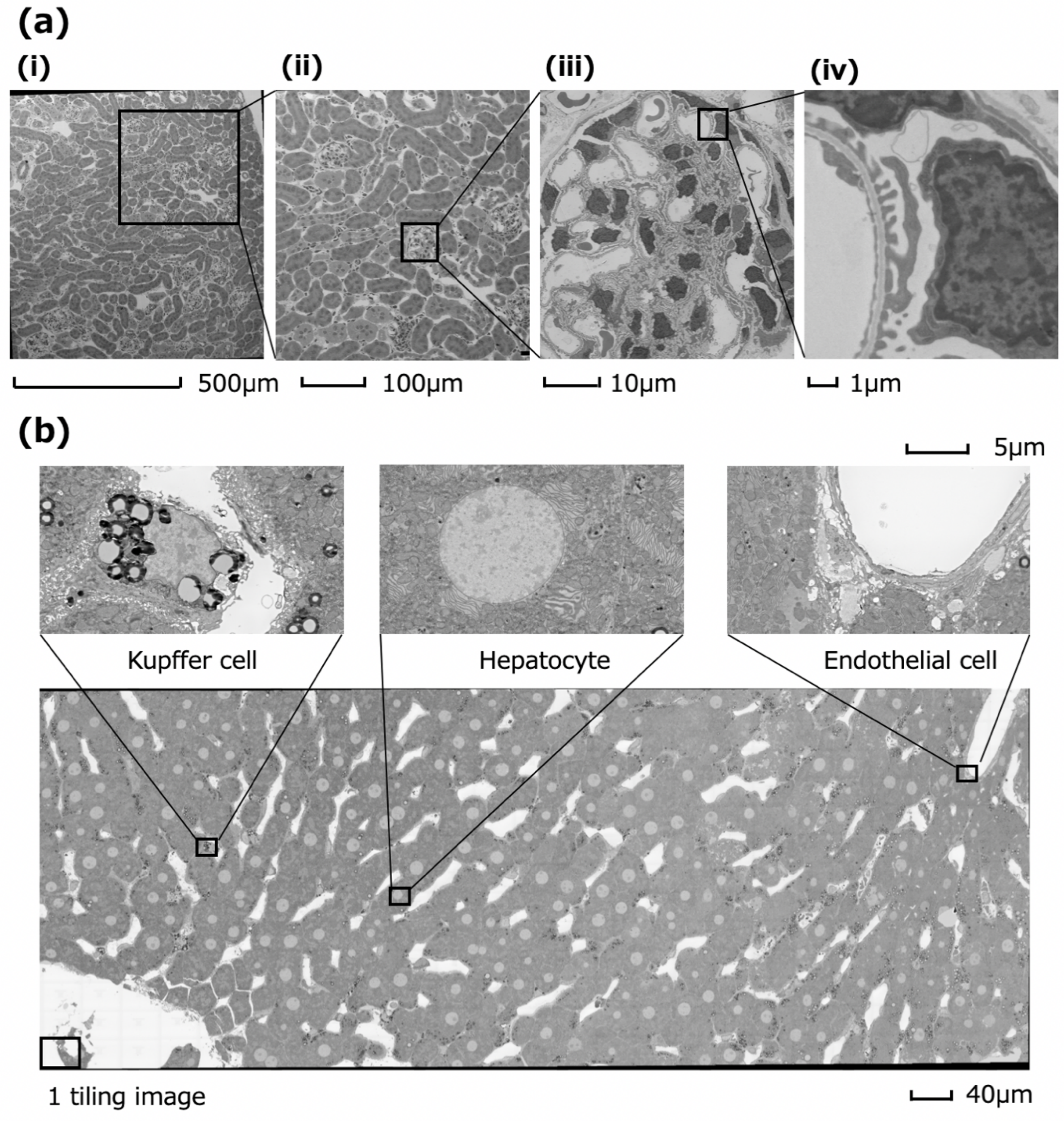

**Fig. 2** 2D wide-area electron microscopy images for the kidney and liver of rodents

(a) Wide-area TEM imaging of a mouse kidney sample. Weakly enlarged images (i and ii) show the simultaneous imaging of multiple glomeruli and renal tubules; strongly magnified images (iii and iv) are conventional EM views containing endothelial basement membranes, podocytes, mesangial cells and epithelial cells. (b) Wide-area SEM imaging of a rat liver section via the backscattered electron detection method. The tiled images provide a view of Kupffer cells, hepatocytes and endothelial cells in addition to the hepatic lobule with different magnifications.

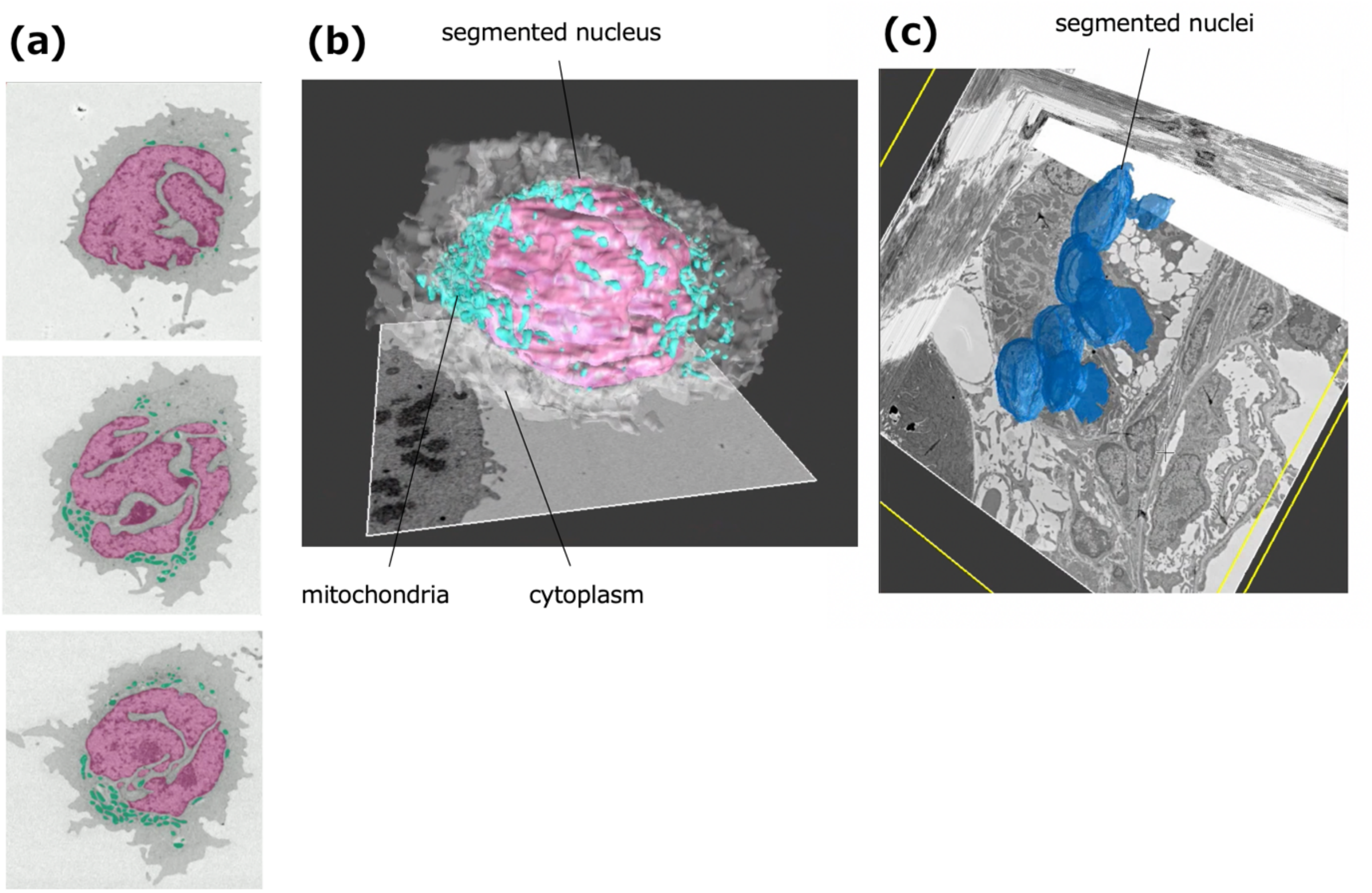

**Fig. 3** 3D volume electron microscopy images obtained using array tomography and SEM

(a) Serial sections of an NB4 cell, i.e. a M3 acute myeloid leukaemia cell line. The shape of the nucleus is highly variable even within a single cell. (b) 3D reconstruction of the NB4 cell using about 130 of the serial sections shown in (a). Nuclei, cell body and mitochondria (high electron density organelles) regions of the cell were segmented. (c) 3D reconstruction of the macula densa in the distal tubules of a mouse kidney glomerulus. Nuclei regions of the macula densa were segmented.

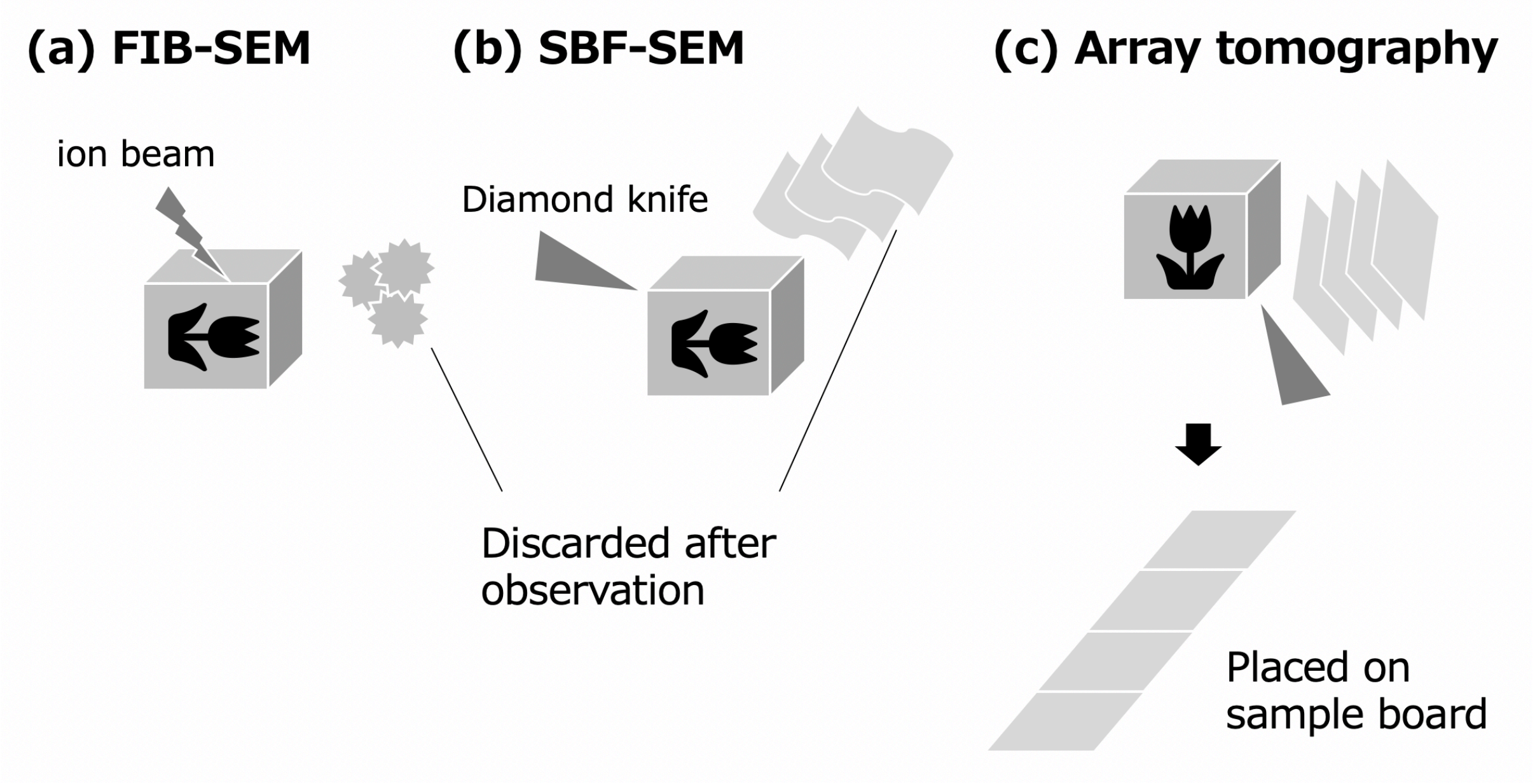

**Fig. 4** The principles of three 3D electron microscopy methods: (a) FIB-SEM, (b) SBF-SEM and (c) array tomography

FIB-SEM and SBF-SEM involve scraping the sample surface with an ion beam and a diamond knife built into the equipment, respectively; they facilitate observation of new cross-sections. In array tomography, serial sections are made in advance and then the same position of each section is observed using EM.

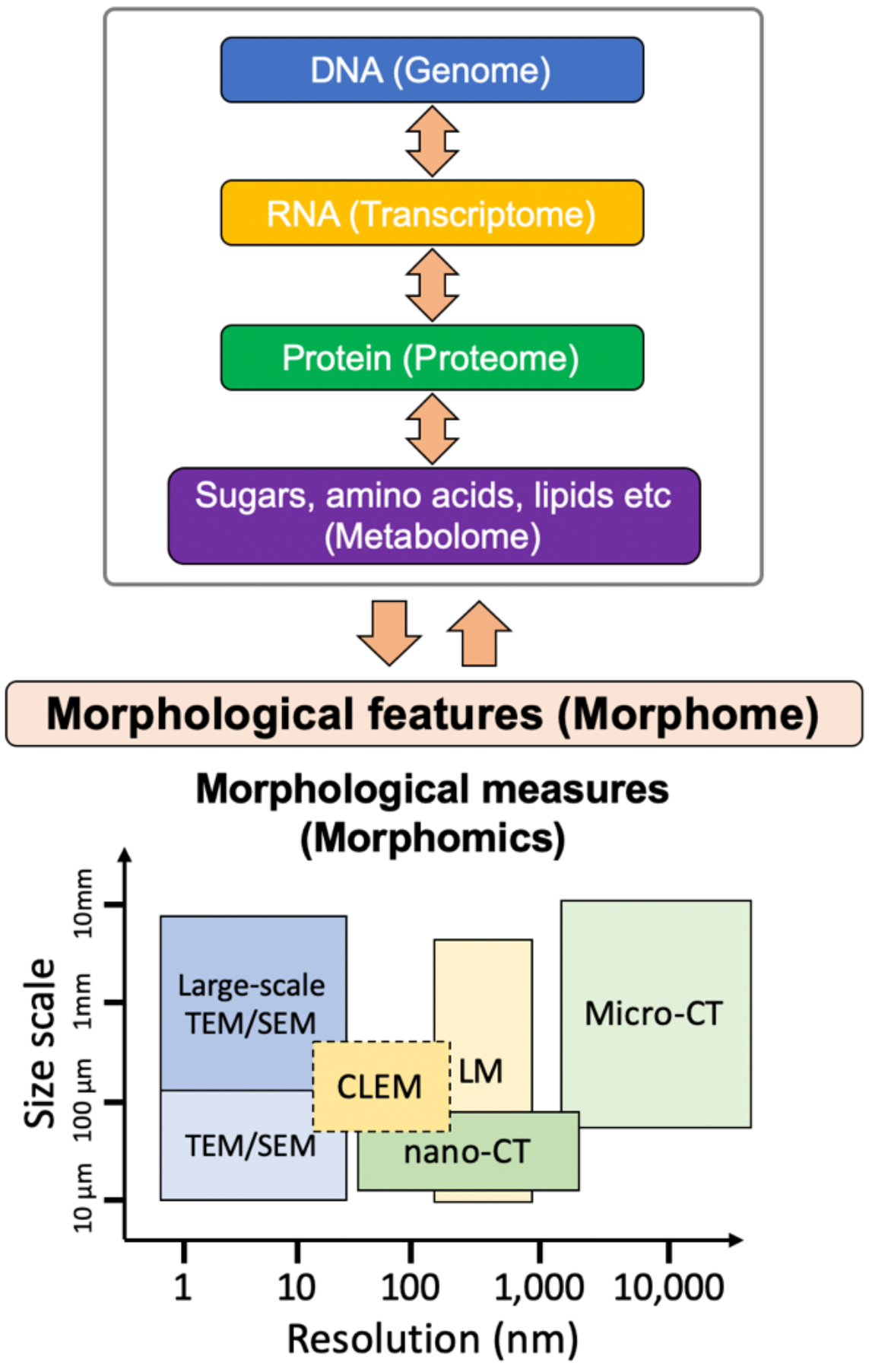

**Fig. 5** Overview of morphomics analysis of biological tissue using various imaging approaches

Biological systems consist of various molecular components such as genes and transcripts (the genome), proteins, lipids, sugars, amino acids and other metabolic components. Quantitative datasets of comprehensive biometric information are usually treated as omics data in biology or bioinformatics. The resulting pattern of molecular information forms the complex structures of tissue and cells in an organism, which is referred to as the morphological features or biological morphome. The nano-scale organisation of the morphome is accessible using the morphomics approaches, such as a large-scale EM, CLEM combined with light microscopy, and CT method, described in this review.

## References

1. Hooke R, Allestry J, Martyn J (1665) Micrographia, or, some physiological descriptions of minute bodies made by magnifying glasses :with observations and inquiries thereupon. London :Printed by Jo. Martyn; Ja. Allestry, printers to the Royal Society ... ,
2. Koch R (1882) Die aetiologie der tuberculose. Berliner Klinische Wochenschrift 15:221–232
3. Noguchi H (1911) METHOD FOR THE PURE CULTIVATION OF PATHOGENIC TREPONEMA PALLIDUM (SPIROCHÆTA PALLIDA) . Journal of Experimental Medicine 14:99–108. https://doi.org/10.1084/jem.14.2.99
4. Leeuwenhoeck M (1674) Microscopical observations from leeuwenhoeck, concerning blood, milk, bones, the brain, spitle, and cuticula, &c. Communicated by the said observer to the publisher in a letter, dated june 1. 1674. Philosophical Transactions of the Royal Society of London 9:121–131. https://doi.org/10.1098/rstl.1674.0030
5. Adelmann HB (1966) Marcello malpighi and the evolution of embryology. Ithaca, NY: Cornell Univ Press
6. Hwa C, Aird WC (2007) The history of the capillary wall: Doctors, discoveries, and debates. American Journal of Physiology-Heart and Circulatory Physiology 293:H2667–H2679. https://doi.org/10.1152/ajpheart.00704.2007
7. Cajal SR (1888) Estructura de los centros nerviosos de las aves. Rev Trim Histol Norm Patol 1:1–10
8. DeFelipe J (2015) The dendritic spine story: An intriguing process of discovery. Frontiers in Neuroanatomy 9:14. https://doi.org/10.3389/fnana.2015.00014
9. López-Muñoz F, Boya J, Alamo C (2006) Neuron theory, the cornerstone of neuroscience, on the centenary of the nobel prize award to santiago ramón y cajal. Brain Research Bulletin 70:391–405. https://doi.org/https://doi.org/10.1016/j.brainresbull.2006.07.010
10. Hall K, Sankaran N (2021) DNA translated: Friedrich miescher's discovery of nuclein in its original context. The British Journal for the History of Science 54:99–107. https://doi.org/10.1017/S000708742000062X
11. Knoll M, Ruska E (1932) Das elektronenmikroskop. Zeitschrift für Physik 78:318–339
12. Ruska E (1987) The development of the electron microscope and of electron microscopy. Bioscience Reports 7:607–629. https://doi.org/10.1007/BF01127674
13. Ackermann H-W (2011) Ruska h. Visualization of bacteriophage lysis in the hypermicroscope. Naturwissenschaften1940; 28:45-6. Bacteriophage 1:183–185. https://doi.org/10.4161/bact.1.4.17624
14. Gelderblom HR, Krüger DH (2014) 1 - helmut ruska (1908–1973): His role in the evolution of electron microscopy in the life sciences, and especially virology. In: Hawkes PW (ed). Elsevier, pp 1–94
15. Richert-Pöggeler KR, Franzke K, Hipp K, Kleespies RG (2019) Electron microscopy methods for virus diagnosis and high resolution analysis of viruses. Frontiers in Microbiology 9:3255. https://doi.org/10.3389/fmicb.2018.03255
16. MARTON L (1934) Electron microscopy of biological objects. Nature 133:911–911. https://doi.org/10.1038/133911b0
17. Brenner S, Horne RW (1959) A negative staining method for high resolution electron microscopy of viruses. Biochimica et Biophysica Acta 34:103–110. https://doi.org/https://doi.org/10.1016/0006-3002(59)90237-9
18. Willingham MC, Rutherford AV (1984) The use of osmium-thiocarbohydrazide-osmium (OTO) and ferrocyanide-reduced osmium methods to enhance membrane contrast and preservation in cultured cells. Journal of Histochemistry & Cytochemistry 32:455–460. https://doi.org/10.1177/32.4.6323574

19. Reynolds ES (1963) THE USE OF LEAD CITRATE AT HIGH pH AS AN ELECTRON-OPAQUE STAIN IN ELECTRON MICROSCOPY. Journal of Cell Biology 17:208–212. https://doi.org/10.1083/jcb.17.1.208
20. HANAICHI T, SATO T, IWAMOTO T, et al (1986) A Stable Lead by Modification of Sato's Method. Journal of Electron Microscopy 35:304–306. https://doi.org/10.1093/oxfordjournals.jmicro.a050582
21. Lewinson D (1989) Application of the ferrocyanide-reduced osmium method for mineralizing cartilage: Further evidence for the enhancement of intracellular glycogen and visualization of matrix components. The Histochemical Journal 21:259–270. https://doi.org/10.1007/BF01757178
22. Pease DC (1964). In: Pease DC (ed) Histological techniques for electron microscopy (second edition), Second Edition. Academic Press
23. Watson ML (1958) Staining of tissue sections for electron microscopy with heavy metals. The Journal of Biophysical and Biochemical Cytology 4:475–478. https://doi.org/10.1083/jcb.4.4.475
24. Stempak JG, Ward RT (1964) AN IMPROVED STAINING METHOD FOR ELECTRON MICROSCOPY . Journal of Cell Biology 22:697–701. https://doi.org/10.1083/jcb.22.3.697
25. Walton J (1979) Lead asparate, an en bloc contrast stain particularly useful for ultrastructural enzymology. Journal of Histochemistry & Cytochemistry 27:1337–1342. https://doi.org/10.1177/27.10.512319
26. Gould RM, Armstrong R (1989) Use of lead aspartate block staining in quantitative EM autoradiography of phospholipids: Application to myelinating peripheral nerve. Journal of Histochemistry & Cytochemistry 37:1393–1399. https://doi.org/10.1177/37.9.2475541
27. OSUMI M (1965) Electron microscopical observations on the formation of yeast-mitochondria. The botanical magazine, Tokyo 78:231–239. https://doi.org/10.15281/jplantres1887.78.231
28. Osumi M, Shimoda C, Yanagishima N (1974) Mating reaction in saccharomyces cerevisiae. Archives of Microbiology 97:27–38. https://doi.org/10.1007/BF00403042
29. Seligman AM, Wasserkrug HL, Hanker JS (1966) A NEW STAINING METHOD (OTO) FOR ENHANCING CONTRAST OF LIPID-CONTAINING MEMBRANES AND DROPLETS IN OSMIUM TETROXIDE-FIXED TISSUE WITH OSMIOPHILIC THIOCARBOHYDRAZIDE (TCH) . Journal of Cell Biology 30:424–432. https://doi.org/10.1083/jcb.30.2.424
30. Cerro MD, Cogen JP, Cerro CD (1981) Retrospective demonstration of endogenous peroxidase activity in plastic-embedded tissues conventionally prepared for electron microscopy. Journal of Histochemistry & Cytochemistry 29:874–876. https://doi.org/10.1177/29.7.7021672
31. Knott G, Genoud C (2013) Is EM dead? Journal of Cell Science 126:4545–4552. https://doi.org/10.1242/jcs.124123
32. Takeshige K, Baba M, Tsuboi S, et al (1992) Autophagy in yeast demonstrated with proteinase-deficient mutants and conditions for its induction. Journal of Cell Biology 119:301–311. https://doi.org/10.1083/jcb.119.2.301
33. Kazimierczak J (1980) A study by scanning (SEM) and transmission (TEM) electron microscopy of the glomerular capillaries in developing rat kidney. Cell and Tissue Research 212:241–255. https://doi.org/10.1007/BF00233959
34. Chen X, Li C, Chen Y, et al (2019) Differentiation of human induced pluripotent stem cells into leydig-like cells with molecular compounds. Cell Death & Disease 10:220. https://doi.org/10.1038/s41419-019-1461-0

35. Ohmine S, Dietz A, Deeds M, et al (2011) Induced pluripotent stem cells from GMP-grade hematopoietic progenitor cells and mononuclear myeloid cells. Stem Cell Research & Therapy 2:46–46
36. Faas FGA, Avramut MC, M. van den Berg B, et al (2012) Virtual nanoscopy: Generation of ultra-large high resolution electron microscopy maps . Journal of Cell Biology 198:457–469. https://doi.org/10.1083/jcb.201201140
37. Bock DD, Lee W-CA, Kerlin AM, et al (2011) Network anatomy and in vivo physiology of visual cortical neurons. Nature 471:177–182. https://doi.org/10.1038/nature09802
38. Kremer JR, Mastronarde DN, McIntosh JR (1996) Computer visualization of three-dimensional image data using IMOD. Journal of Structural Biology 116:71–76. https://doi.org/https://doi.org/10.1006/jsbi.1996.0013
39. Titze B, Genoud C (2016) Volume scanning electron microscopy for imaging biological ultrastructure. Biology of the Cell 108:307–323. https://doi.org/https://doi.org/10.1111/boc.201600024
40. Wacker I, Spomer W, Hofmann A, et al (2016) Hierarchical imaging: A new concept for targeted imaging of large volumes from cells to tissues. BMC Cell Biology 17:38. https://doi.org/10.1186/s12860-016-0122-8
41. Schalek R, Kasthuri N, Hayworth K, et al (2011) Development of high-throughput, high-resolution 3D reconstruction of large-volume biological tissue using automated tape collection ultramicrotomy and scanning electron microscopy. Microscopy and Microanalysis 17:966–967. https://doi.org/10.1017/S1431927611005708
42. Kume S, Masuya H, Maeda M, et al (2017) Development of semantic web-based imaging database for biological morphome. In: Wang Z, Turhan A-Y, Wang K, Zhang X (eds) Semantic technology. Springer International Publishing, Cham, pp 277–285
43. Baruzzi A, Caveggion E, Berton G (2008) Regulation of phagocyte migration and recruitment by src-family kinases. Cellular and Molecular Life Sciences 65:2175–2190. https://doi.org/10.1007/s00018-008-8005-6
44. Mironov AA, Beznoussenko GV (2019) Models of intracellular transport: Pros and cons. Frontiers in Cell and Developmental Biology 7:146. https://doi.org/10.3389/fcell.2019.00146
45. O'Connor CM, Adams JU (2010) Essentials of cell biology. Cambridge, MA: NPG Education
46. Casares D, Escribá PV, Rosselló CA (2019) Membrane lipid composition: Effect on membrane and organelle structure, function and compartmentalization and therapeutic avenues. International Journal of Molecular Sciences 20: https://doi.org/10.3390/ijms20092167
47. GLAUERT AM, ROGERS GE, GLAUERT RH (1956) A new embedding medium for electron microscopy. Nature 178:803–803. https://doi.org/10.1038/178803a0
48. Glauert AM, Glauert RH (1958) Araldite as an embedding medium for electron microscopy. The Journal of Biophysical and Biochemical Cytology 4:191–194. https://doi.org/10.1083/jcb.4.2.191
49. Wigglesworth VB (1957) The use of osmium in the fixation and staining of tissues. Proceedings of the Royal Society of London Series B - Biological Sciences 147:185–199. https://doi.org/10.1098/rspb.1957.0043
50. Maeda M, Seto T, Kadono C, et al (2019) Autophagy in the central nervous system and effects of chloroquine in mucopolysaccharidosis type II mice. International Journal of Molecular Sciences 20: https://doi.org/10.3390/ijms20235829
51. Kiernan JA (2007) Histochemistry of staining methods for normal and degenerating myelin in the central and peripheral nervous systems. Journal of Histotechnology 30:87–106. https://doi.org/10.1179/his.2007.30.2.87

52. Hua Y, Laserstein P, Helmstaedter M (2015) Large-volume en-bloc staining for electron microscopy-based connectomics. Nature Communications 6:7923. https://doi.org/10.1038/ncomms8923
53. Watson ML (1958) Staining of Tissue Sections for Electron Microscopy with Heavy Metals : II. Application of Solutions Containing Lead and Barium . The Journal of Biophysical and Biochemical Cytology 4:727–730. https://doi.org/10.1083/jcb.4.6.727
54. Nakakoshi M, Nishioka H, Katayama E (2011) New versatile staining reagents for biological transmission electron microscopy that substitute for uranyl acetate. Journal of Electron Microscopy 60:401–407. https://doi.org/10.1093/jmicro/dfr084
55. Kuipers J, Giepmans BNG (2020) Neodymium as an alternative contrast for uranium in electron microscopy. Histochemistry and Cell Biology 153:271–277. https://doi.org/10.1007/s00418-020-01846-0
56. Mikula S, Denk W (2015) High-resolution whole-brain staining for electron microscopic circuit reconstruction. Nature Methods 12:541–546. https://doi.org/10.1038/nmeth.3361
57. Hua Y, Ding X, Wang H, et al (2021) Electron microscopic reconstruction of neural circuitry in the cochlea. Cell Reports 34:108551. https://doi.org/https://doi.org/10.1016/j.celrep.2020.108551
58. Mikula S, Binding J, Denk W (2012) Staining and embedding the whole mouse brain for electron microscopy. Nature Methods 9:1198–1201. https://doi.org/10.1038/nmeth.2213
59. Deerinck T, Bushong E, Lev-Ram V, et al (2010) Enhancing serial block-face scanning electron microscopy to enable high resolution 3-d nanohistology of cells and tissues. Microscopy and Microanalysis 16:1138–1139. https://doi.org/10.1017/S1431927610055170
60. Polilov AA, Makarova AA, Pang S, et al (2021) Protocol for preparation of heterogeneous biological samples for 3D electron microscopy: A case study for insects. Scientific Reports 11:4717. https://doi.org/10.1038/s41598-021-83936-0
61. Raméntol NC i, García A, Fernández E (2012) Transmission electron microscopy in cell biology: Sample preparation techniques and image information
62. Sant' Agnese PAD, De Mesy Jensen KL (1984) Dibasic Staining of Large Epoxy Tissue Sections and Applications to Surgical Pathology. American Journal of Clinical Pathology 81:25–29. https://doi.org/10.1093/ajcp/81.1.25
63. Luft JH (1961) IMPROVEMENTS IN EPOXY RESIN EMBEDDING METHODS . The Journal of Biophysical and Biochemical Cytology 9:409–414. https://doi.org/10.1083/jcb.9.2.409
64. Trump BF, Smuckler EA, Benditt EP (1961) A method for staining epoxy sections for light microscopy. Journal of Ultrastructure Research 5:343–348. https://doi.org/https://doi.org/10.1016/S0022-5320(61)80011-7
65. Nagashima K, Zheng J, Parmiter D, Patri AK (2011) Biological tissue and cell culture specimen preparation for TEM nanoparticle characterization. In: McNeil SE (ed) Characterization of nanoparticles intended for drug delivery. Humana Press, Totowa, NJ, pp 83–91
66. Ardenne MV, Wissensch Z (1939). Mikroskope 56:8
67. Pease DC, Baker RF (19481948) Sectioning techniques for electron microscopy using a conventional microtome. Proceedings of the Society for Experimental Biology and Medicine 67:470–474. https://doi.org/10.3181/00379727-67-16344
68. Sjöstrand FS (1953) The ultrastructure of the inner segments of the retinal rods of the guinea pig eye as revealed by electron microscopy. Journal of Cellular and Comparative Physiology 42:45–70. https://doi.org/https://doi.org/10.1002/jcp.1030420104
69. Porter KR, Blum J (1953) A study in microtomy for electron microscopy. The Anatomical Record 117:685–709. https://doi.org/https://doi.org/10.1002/ar.1091170403

70. Fernández-Morán H (1953) A diamond knife for ultrathin sectioning. Experimental Cell Research 5:255–256. https://doi.org/https://doi.org/10.1016/0014-4827(53)90112-8
71. Fernández-Morán H (1956) Fine structure of the insect retinula as revealed by electron microscopy. Nature 177:742–743. https://doi.org/10.1038/177742a0
72. Fernández-Morán H, Oda T, Blair PV, Green DE (1964) A MACROMOLECULAR REPEATING UNIT OF MITOCHONDRIAL STRUCTURE AND FUNCTION : Correlated Electron Microscopic and Biochemical Studies of Isolated Mitochondria and Submitochondrial Particles of Beef Heart Muscle . Journal of Cell Biology 22:63–100. https://doi.org/10.1083/jcb.22.1.63
73. Plummer HK (1997) Reflections on the use of microtomy for materials science specimen preparation. Microscopy and Microanalysis 3:239–260. https://doi.org/10.1017/S1431927697970197
74. Xu Q, Rioux RM, Whitesides GM (2007) Fabrication of complex metallic nanostructures by nanoskiving. ACS Nano 1:215–227. https://doi.org/10.1021/nn700172c
75. Jésior J-C (1986) How to avoid compression II. The influence of sectioning conditions. Journal of Ultrastructure and Molecular Structure Research 95:210–217. https://doi.org/https://doi.org/10.1016/0889-1605(86)90042-X
76. Sun F, Li H, Leifer K, Gamstedt EK (2017) Rate effects on localized shear deformation during nanosectioning of an amorphous thermoplastic polymer. International Journal of Solids and Structures 129:40–48. https://doi.org/https://doi.org/10.1016/j.ijsolstr.2017.09.016
77. Gay H, Anderson TF (1954) Serial sections for electron microscopy. Science 120:1071–1073. https://doi.org/10.1126/science.120.3130.1071
78. Miranda K, Girard-Dias W, Attias M, et al (2015) Three dimensional reconstruction by electron microscopy in the life sciences: An introduction for cell and tissue biologists. Molecular Reproduction and Development 82:530–547. https://doi.org/https://doi.org/10.1002/mrd.22455
79. Hayworth KJ, Morgan JL, Schalek R, et al (2014) Imaging ATUM ultrathin section libraries with WaferMapper: A multi-scale approach to EM reconstruction of neural circuits. Frontiers in Neural Circuits 8:68. https://doi.org/10.3389/fncir.2014.00068
80. Baena V, Schalek RL, Lichtman JW, Terasaki M (2019) Chapter 3 - serial-section electron microscopy using automated tape-collecting ultramicrotome (ATUM). In: Müller-Reichert T, Pigino G (eds) Three-dimensional electron microscopy. Academic Press, pp 41–67
81. Kasthuri N, Hayworth KJ, Berger DR, et al (2015) Saturated reconstruction of a volume of neocortex. Cell 162:648–661. https://doi.org/10.1016/j.cell.2015.06.054
82. Kubota Y, Sohn J, Hatada S, et al (2018) A carbon nanotube tape for serial-section electron microscopy of brain ultrastructure. Nature Communications 9:437. https://doi.org/10.1038/s41467-017-02768-7
83. Burel A, Lavault M-T, Chevalier C, et al (2018) A targeted 3D EM and correlative microscopy method using SEM array tomography. Development 145: https://doi.org/10.1242/dev.160879
84. Winey M, Meehl JB, O'Toole ET, Giddings TH (2014) Conventional transmission electron microscopy. Molecular Biology of the Cell 25:319–323. https://doi.org/10.1091/mbc.e12-12-0863
85. Graham L, Orenstein JM (2007) Processing tissue and cells for transmission electron microscopy in diagnostic pathology and research. Nature Protocols 2:2439–2450. https://doi.org/10.1038/nprot.2007.304
86. KRUMEICH F (2018) INTRODUCTION INTO TRANSMISSION AND SCANNING TRANSMISSION ELECTRON MICROSCOPY

87. Krumeich F (2015) Properties of electrons, their interactions with matter and applications in electron microscopy
88. Ravelli RBG, Kalicharan RD, Avramut MC, et al (2013) Destruction of tissue, cells and organelles in type 1 diabetic rats presented at macromolecular resolution. Scientific Reports 3:1804. https://doi.org/10.1038/srep01804
89. Al-Amoudi A, Studer D, Dubochet J (2005) Cutting artefacts and cutting process in vitreous sections for cryo-electron microscopy. Journal of Structural Biology 150:109–121. https://doi.org/https://doi.org/10.1016/j.jsb.2005.01.003
90. Preibisch S, Saalfeld S, Tomancak P (2009) Globally optimal stitching of tiled 3D microscopic image acquisitions. Bioinformatics 25:1463–1465. https://doi.org/10.1093/bioinformatics/btp184
91. Toyooka K, Sato M, Kutsuna N, et al (2014) Wide-Range High-Resolution Transmission Electron Microscopy Reveals Morphological and Distributional Changes of Endomembrane Compartments during Log to Stationary Transition of Growth Phase in Tobacco BY-2 Cells. Plant and Cell Physiology 55:1544–1555. https://doi.org/10.1093/pcp/pcu084
92. Gordon RE (2014) Electron microscopy: A brief history and review of current clinical application. In: Day CE (ed) Histopathology: Methods and protocols. Springer New York, New York, NY, pp 119–135
93. Lichtman JW, Denk W (2011) The big and the small: Challenges of imaging the brain's circuits. Science 334:618–623. https://doi.org/10.1126/science.1209168
94. Quan TM, Hildebrand DGC, Lee K, et al (2019) Removing imaging artifacts in electron microscopy using an asymmetrically cyclic adversarial network without paired training data. In: 2019 IEEE/CVF international conference on computer vision workshop (ICCVW). pp 3804–3813
95. Konyuba Y, Haruta T, Ikeda Y, Fukuda T (2018) Fabrication and characterization of sample-supporting film made of silicon nitride for large-area observation in transmission electron microscopy. Microscopy 67:367–370. https://doi.org/10.1093/jmicro/dfy039
96. Lefman J, Morrison R, Subramaniam S (2007) Automated 100-position specimen loader and image acquisition system for transmission electron microscopy. Journal of Structural Biology 158:318–326. https://doi.org/https://doi.org/10.1016/j.jsb.2006.11.007
97. Schorb M, Haberbosch I, Hagen WJH, et al (2019) Software tools for automated transmission electron microscopy. Nature Methods 16:471–477. https://doi.org/10.1038/s41592-019-0396-9
98. Kaynig V, Fischer B, Müller E, Buhmann JM (2010) Fully automatic stitching and distortion correction of transmission electron microscope images. Journal of Structural Biology 171:163–173. https://doi.org/https://doi.org/10.1016/j.jsb.2010.04.012
99. Cardona A, Saalfeld S, Schindelin J, et al (2012) TrakEM2 software for neural circuit reconstruction. PLOS ONE 7:1–8. https://doi.org/10.1371/journal.pone.0038011
100. Pereira AF, Hageman DJ, Garbowski T, et al (2016) Creating high-resolution multiscale maps of human tissue using multi-beam SEM. PLOS Computational Biology 12:1–17. https://doi.org/10.1371/journal.pcbi.1005217
101. Buckman J (2014) Use of automated image acquisition and stitching in scanning electron microscopy: Imaging of large scale areas of materials at high resolution: SEM image stitching. Microscopy and Analysis January 2014:s13–16
102. Ma B, Zimmermann T, Rohde M, et al (2007) Use of autostitch for automatic stitching of microscope images. Micron 38:492–499. https://doi.org/https://doi.org/10.1016/j.micron.2006.07.027

103. Chalfoun J, Majurski M, Blattner T, et al (2017) MIST: Accurate and scalable microscopy image stitching tool with stage modeling and error minimization. Scientific Reports 7:4988. https://doi.org/10.1038/s41598-017-04567-y
104. Blattner T, Keyrouz W, Chalfoun J, et al (2014) A hybrid CPU-GPU system for stitching large scale optical microscopy images. In: 2014 43rd international conference on parallel processing. pp 1–9
105. Saalfeld S (2019) Chapter 12 - computational methods for stitching, alignment, and artifact correction of serial section data. In: Müller-Reichert T, Pigino G (eds) Three-dimensional electron microscopy. Academic Press, pp 261–276
106. TSAI C-L, LISTER JP, BJORNSSON CS, et al (2011) Robust, globally consistent and fully automatic multi-image registration and montage synthesis for 3-d multi-channel images. Journal of Microscopy 243:154–171. https://doi.org/https://doi.org/10.1111/j.1365-2818.2011.03489.x
107. Liu D, Wang S, Cao P, et al (2013) Dark-field microscopic image stitching method for surface defects evaluation of large fine optics. Opt Express 21:5974–5987. https://doi.org/10.1364/OE.21.005974
108. Higaki T, Kutsuna N, Akita K, et al (2015) Semi-automatic organelle detection on transmission electron microscopic images. Scientific Reports 5:7794. https://doi.org/10.1038/srep07794
109. Toyooka K, Sato M, Wakazaki M, Matsuoka K (2016) Morphological and quantitative changes in mitochondria, plastids, and peroxisomes during the log-to-stationary transition of the growth phase in cultured tobacco BY-2 cells. Plant Signaling & Behavior 11:e1149669. https://doi.org/10.1080/15592324.2016.1149669
110. Lamers MM, Beumer J, Vaart J van der, et al (2020) SARS-CoV-2 productively infects human gut enterocytes. Science 369:50–54. https://doi.org/10.1126/science.abc1669
111. AU - Trépout S, AU - Bastin P, AU - Marco S (2017) Preparation and observation of thick biological samples by scanning transmission electron tomography. JoVE e55215. https://doi.org/10.3791/55215
112. Boer P de, Pirozzi NM, Wolters AHG, et al (2020) Large-scale electron microscopy database for human type 1 diabetes. Nature Communications 11:2475. https://doi.org/10.1038/s41467-020-16287-5
113. Sokol E, Kramer D, Diercks GFH, et al (2015) Large-scale electron microscopy maps of patient skin and mucosa provide insight into pathogenesis of blistering diseases. Journal of Investigative Dermatology 135:1763–1770. https://doi.org/https://doi.org/10.1038/jid.2015.109
114. Siegert E, Uruha A, Goebel H-H, et al (2021) Systemic sclerosis-associated myositis features minimal inflammation and characteristic capillary pathology. Acta Neuropathologica 141:917–927. https://doi.org/10.1007/s00401-021-02305-3
115. Dittmayer C, Meinhardt J, Radbruch H, et al (2020) Why misinterpretation of electron micrographs in SARS-CoV-2-infected tissue goes viral. The Lancet 396:e64–e65. https://doi.org/10.1016/S0140-6736(20)32079-1
116. Capala ME, Maat H, Bonardi F, et al (2015) Mitochondrial dysfunction in human leukemic stem/progenitor cells upon loss of RAC2. PLOS ONE 10:1–20. https://doi.org/10.1371/journal.pone.0128585
117. AU - Kuipers J, AU - Kalicharan RD, AU - Wolters AHG, et al (2016) Large-scale scanning transmission electron microscopy (nanotomy) of healthy and injured zebrafish brain. JoVE e53635. https://doi.org/10.3791/53635
118. Grudniewska M, Mouton S, Grelling M, et al (2018) A novel flatworm-specific gene implicated in reproduction in macrostomum lignano. Scientific Reports 8:3192. https://doi.org/10.1038/s41598-018-21107-4

119. Kuipers J, de Boer P, Giepmans BNG (2015) Scanning EM of non-heavy metal stained biosamples: Large-field of view, high contrast and highly efficient immunolabeling. Experimental Cell Research 337:202–207. https://doi.org/https://doi.org/10.1016/j.yexcr.2015.07.012
120. Scotuzzi M, Kuipers J, Wensveen DI, et al (2017) Multi-color electron microscopy by element-guided identification of cells, organelles and molecules. Scientific Reports 7:45970. https://doi.org/10.1038/srep45970
121. Vos P de, Smink AM, Paredes G, et al (2016) Enzymes for pancreatic islet isolation impact chemokine-production and polarization of insulin-producing β-cells with reduced functional survival of immunoisolated rat islet-allografts as a consequence. PLOS ONE 11:1–18. https://doi.org/10.1371/journal.pone.0147992
122. Maimets M, Rocchi C, Bron R, et al (2016) Long-term in vitro expansion of salivary gland stem cells driven by wnt signals. Stem Cell Reports 6:150–162. https://doi.org/10.1016/j.stemcr.2015.11.009
123. Nijholt KT, Meems LMG, Ruifrok WPT, et al (2021) The erythropoietin receptor expressed in skeletal muscle is essential for mitochondrial biogenesis and physiological exercise. Pflügers Archiv - European Journal of Physiology 473:1301–1313. https://doi.org/10.1007/s00424-021-02577-4
124. Dittmayer C, Völcker E, Wacker I, et al (2018) Modern field emission scanning electron microscopy provides new perspectives for imaging kidney ultrastructure. Kidney International 94:625–631. https://doi.org/https://doi.org/10.1016/j.kint.2018.05.017
125. Dane MJC, Berg BM van den, Avramut MC, et al (2013) Glomerular endothelial surface layer acts as a barrier against albumin filtration. The American Journal of Pathology 182:1532–1540. https://doi.org/10.1016/j.ajpath.2013.01.049
126. Vrij EL de, Bouma HR, Goris M, et al (2021) Reversible thrombocytopenia during hibernation originates from storage and release of platelets in liver sinusoids. Journal of Comparative Physiology B 191:603–615. https://doi.org/10.1007/s00360-021-01351-3
127. Dittmayer C, Goebel H-H, Heppner FL, et al (2021) Preparation of samples for large-scale automated electron microscopy of tissue and cell ultrastructure. Microscopy and Microanalysis 27:815–827. https://doi.org/10.1017/S1431927621011958
128. Pirozzi NM, Hoogenboom JP, Giepmans BNG (2018) ColorEM: Analytical electron microscopy for element-guided identification and imaging of the building blocks of life. Histochemistry and Cell Biology 150:509–520. https://doi.org/10.1007/s00418-018-1707-4
129. Pirozzi NM, Kuipers J, Giepmans BNG (2021) Chapter 5 - sample preparation for energy dispersive x-ray imaging of biological tissues. In: Müller-Reichert T, Verkade P (eds) Correlative light and electron microscopy IV. Academic Press, pp 89–114
130. McMullan D (2008) The early development of the scanning electron microscope. In: Schatten H, Pawley JB (eds) Biological low-voltage scanning electron microscopy. Springer New York, New York, NY, pp 1–25
131. Bogner A, Jouneau P-H, Thollet G, et al (2007) A history of scanning electron microscopy developments: Towards “wet-STEM” imaging. Micron 38:390–401. https://doi.org/https://doi.org/10.1016/j.micron.2006.06.008
132. Suga M, Asahina S, Sakuda Y, et al (2014) Recent progress in scanning electron microscopy for the characterization of fine structural details of nano materials. Progress in Solid State Chemistry 42:1–21. https://doi.org/https://doi.org/10.1016/j.progsolidstchem.2014.02.001
133. Bouwer JC, Deerinck TJ, Bushong E, et al (2016) Deceleration of probe beam by stage bias potential improves resolution of serial block-face scanning electron microscopic images. Advanced Structural and Chemical Imaging 2:11. https://doi.org/10.1186/s40679-016-0025-y
134. Scala C, Cenacchi G, Preda P, et al (1991) Conventional and high resolution scanning electron microscopy of biological sectioned material. Scanning microscopy 5 1:135–145
135. Rodríguez J-R, Turégano-López M, DeFelipe J, Merchán-Pérez A (2018) Neuroanatomy from mesoscopic to nanoscopic scales: An improved method for the observation of semithin sections by high-resolution scanning electron microscopy. Frontiers in Neuroanatomy 12:14. https://doi.org/10.3389/fnana.2018.00014
136. Walther P, Autrata R, Chen Y, Pawley J (1991) Backscattered electron imaging for high resolution surface scanning electron microscopy with a new type YAG-detector. Scanning microscopy 5 2:301–9; discussion 310
137. Conti S, Perico N, Novelli R, et al (2018) Early and late scanning electron microscopy findings in diabetic kidney disease. Scientific Reports 8:4909. https://doi.org/10.1038/s41598-018-23244-2
138. Autio-Harmainen H, Väänänen R, Rapola J (1981) Scanning electron microscopic study of normal human glomerulogenesis and of fetal glomeruli in congenital nephrotic syndrome of the finnish type. Kidney International 20:747–752. https://doi.org/10.1038/ki.1981.206
139. Burghardt T, Hochapfel F, Salecker B, et al (2015) Advanced electron microscopic techniques provide a deeper insight into the peculiar features of podocytes. American Journal of Physiology-Renal Physiology 309:F1082–F1089. https://doi.org/10.1152/ajprenal.00338.2015
140. Bonsib SM (1985) Scanning electron microscopy of acellular glomeruli in nephrotic syndrome. Kidney International 27:678–684. https://doi.org/10.1038/ki.1985.64
141. Bonsib SM (1988) Glomerular basement membrane necrosis and crescent organization. Kidney International 33:966–974. https://doi.org/10.1038/ki.1988.95
142. Grahammer F, Wigge C, Schell C, et al (2017) A flexible, multilayered protein scaffold maintains the slit in between glomerular podocytes. JCI Insight 1: https://doi.org/10.1172/jci.insight.86177
143. De Cesare F, Di Mattia E, Zussman E, Macagnano A (2019) A study on the dependence of bacteria adhesion on the polymer nanofibre diameter. Environ Sci: Nano 6:778–797. https://doi.org/10.1039/C8EN01237G
144. Khalaf S, Ariffin Z, Husein A, Reza F (2017) Surface coating of gypsum-based molds for maxillofacial prosthetic silicone elastomeric material: Evaluating different microbial adhesion. Journal of prosthodontics : official journal of the American College of Prosthodontists 26:664—669. https://doi.org/10.1111/jopr.12460
145. Boatman E, Cartwright F, Kenny G (1976) Morphology, morphometry and electron microscopy of HeLa cells infected with bovine mycoplasma. Cell and tissue research 170 1:1–16
146. Caldas LA, Carneiro FA, Higa LM, et al (2020) Ultrastructural analysis of SARS-CoV-2 interactions with the host cell via high resolution scanning electron microscopy. Scientific Reports 10:16099. https://doi.org/10.1038/s41598-020-73162-5
147. Tokunaga J, Fujita T, Hattori A (1969) Scanning electron microscopy of normal and pathological human erythrocytes. Archivum histologicum Japonicum = Nihon soshikigaku kiroku 31(1):21–35
148. Buys AV, Van Rooy M-J, Soma P, et al (2013) Changes in red blood cell membrane structure in type 2 diabetes: A scanning electron and atomic force microscopy study. Cardiovascular Diabetology 12:25. https://doi.org/10.1186/1475-2840-12-25
149. Hattori A, Ito S, Matsuoka M (1972) Scanning electron microscopy of reticulocytes. Archivum histologicum Japonicum = Nihon soshikigaku kiroku 35 1:37–49

150. Grindem CB (1985) Ultrastructural morphology of leukemic cells in the cat. Veterinary Pathology 22:147–155. https://doi.org/10.1177/030098588502200209
151. Hartigan A, Estensoro I, Vancová M, et al (2016) New cell motility model observed in parasitic cnidarian sphaerospora molnari (myxozoa:myxosporea) blood stages in fish. Scientific Reports 6:39093. https://doi.org/10.1038/srep39093
152. Cohen Hyams T, Mam K, Killingsworth MC (2020) Scanning electron microscopy as a new tool for diagnostic pathology and cell biology. Micron 130:102797. https://doi.org/https://doi.org/10.1016/j.micron.2019.102797
153. Joy DC (1991) The theory and practice of high-resolution scanning electron microscopy. Ultramicroscopy 37:216–233. https://doi.org/https://doi.org/10.1016/0304-3991(91)90020-7
154. Dittmayer C, Völcker E, Wacker I, et al (2018) Modern field emission scanning electron microscopy provides new perspectives for imaging kidney ultrastructure. Kidney International 94:625–631. https://doi.org/10.1016/j.kint.2018.05.017
155. RICHARDS RG, GWYNN IA (1995) Backscattered electron imaging of the undersurface of resin-embedded cells by field-emission scanning electron microscopy. Journal of Microscopy 177:43–52. https://doi.org/https://doi.org/10.1111/j.1365-2818.1995.tb03532.x
156. Koga D, Kusumi S, Watanabe T (2018) Backscattered electron imaging of resin-embedded sections. Microscopy 67:196–206. https://doi.org/10.1093/jmicro/dfy028
157. Crewe AV, Eggenberger DN, Wall J, Welter LM (1968) Electron gun using a field emission source. Review of Scientific Instruments 39:576–583. https://doi.org/10.1063/1.1683435
158. Swanson LW, Crouser LC (1969) Angular confinement of field electron and ion emission. Journal of Applied Physics 40:4741–4749. https://doi.org/10.1063/1.1657282
159. Swanson L, Schwind G (2008) Review of ZrO/w schottky cathode in: Orloff jon, editor. Handbook of charged particle optics. 2nd ed. CRC press. pp 1–28
160. Williams TJ (2005) Scanning electron microscopy and x-ray microanalysis, 3rd edition. By joseph goldstein, dale newbury, david joy, charles lyman, patrick echlin, eric lifshin, linda sawyer, joseph michael kluwer academic publishers, new york (2003). Scanning 27:215–216. https://doi.org/https://doi.org/10.1002/sca.4950270410
161. Lane R, Vos Y, Wolters AHG, et al (2021) Optimization of negative stage bias potential for faster imaging in large-scale electron microscopy. Journal of Structural Biology: X 5:100046. https://doi.org/https://doi.org/10.1016/j.yjsbx.2021.100046
162. Reimer L, Riepenhausen M (1985) Detector strategy for secondary and backscattered electrons using multiple detector systems. Scanning 7:221–238. https://doi.org/https://doi.org/10.1002/sca.4950070503
163. Borzunov AA, Karaulov VY, Koshev NA, et al (2019) 3D surface topography imaging in SEM with improved backscattered electron detector: Arrangement and reconstruction algorithm. Ultramicroscopy 207:112830. https://doi.org/https://doi.org/10.1016/j.ultramic.2019.112830
164. Knott G, Marchman H, Wall D, Lich B (2008) Serial section scanning electron microscopy of adult brain tissue using focused ion beam milling. Journal of Neuroscience 28:2959–2964. https://doi.org/10.1523/JNEUROSCI.3189-07.2008
165. Brahim Belhaouari D, Fontanini A, Baudoin J-P, et al (2020) The strengths of scanning electron microscopy in deciphering SARS-CoV-2 infectious cycle. Frontiers in Microbiology 11:2014. https://doi.org/10.3389/fmicb.2020.02014
166. Koster AJ, Klumperman J (2003) Electron microscopy in cell biology: Integrating structure and function. Nature reviews Molecular cell biology Suppl:SS6—10
167. Bozzola JJ (2002) Electron microscopy. In: Encyclopedia of Life Sciences 1–10

168. Satir P (2005) Tour of organelles through the electron microscope: A reprinting of keith r. Porter's classic harvey lecture with a new introduction. The Anatomical Record Part A: Discoveries in Molecular, Cellular, and Evolutionary Biology 287A:1184–1204. https://doi.org/https://doi.org/10.1002/ar.a.20222
169. Bainton DF, Friedlander LM, Shohet SB (1977) Abnormalities in granule formation in acute myelogenous leukemia. Blood 49:693–704. https://doi.org/https://doi.org/10.1182/blood.V49.5.693.693
170. Miyauchi J, Ohyashiki K, Inatomi Y, Toyama K (1997) Neutrophil Secondary-Granule Deficiency as a Hallmark of All-Trans Retinoic Acid–Induced Differentiation of Acute Promyelocytic Leukemia Cells. Blood 90:803–813. https://doi.org/10.1182/blood.V90.2.803
171. Naito M, Hasegawa G, Ebe Y, Yamamoto T (2004) Differentiation and function of kupffer cells. Medical electron microscopy : official journal of the Clinical Electron Microscopy Society of Japan 37:16—28. https://doi.org/10.1007/s00795-003-0228-x
172. Ioannou GN, Haigh WG, Thorning D, Savard C (2013) Hepatic cholesterol crystals and crown-like structures distinguish NASH from simple steatosis [s]. Journal of Lipid Research 54:1326–1334. https://doi.org/10.1194/jlr.M034876
173. Matey V, Richards JG, Wang Y, et al (2008) The effect of hypoxia on gill morphology and ionoregulatory status in the Lake Qinghai scaleless carp, Gymnocypris przewalskii. Journal of Experimental Biology 211:1063–1074. https://doi.org/10.1242/jeb.010181
174. Rice WL, Van Hoek AN, Păunescu TG, et al (2013) High resolution helium ion scanning microscopy of the rat kidney. PLOS ONE 8:1–9. https://doi.org/10.1371/journal.pone.0057051
175. Păunescu TG, Shum WWC, Huynh C, et al (2014) High-resolution helium ion microscopy of epididymal epithelial cells and their interaction with spermatozoa. Molecular Human Reproduction 20:929–937. https://doi.org/10.1093/molehr/gau052
176. Tsuji K, Suleiman H, Miner JH, et al (2017) Ultrastructural characterization of the glomerulopathy in alport mice by helium ion scanning microscopy (HIM). Scientific Reports 7:11696. https://doi.org/10.1038/s41598-017-12064-5
177. Tsuji K, Păunescu TG, Suleiman H, et al (2017) Re-characterization of the glomerulopathy in CD2AP deficient mice by high-resolution helium ion scanning microscopy. Scientific Reports 7:8321. https://doi.org/10.1038/s41598-017-08304-3
178. GOLLA U, SCHINDLER B, REIMER L (1994) Contrast in the transmission mode of a low-voltage scanning electron microscope. Journal of Microscopy 173:219–225. https://doi.org/https://doi.org/10.1111/j.1365-2818.1994.tb03444.x
179. Kuwajima M, Mendenhall JM, Lindsey LF, Harris KM (2013) Automated transmission-mode scanning electron microscopy (tSEM) for large volume analysis at nanoscale resolution. PLOS ONE 8:1–14. https://doi.org/10.1371/journal.pone.0059573
180. Suga M, Nishiyama H, Konyuba Y, et al (2011) The atmospheric scanning electron microscope with open sample space observes dynamic phenomena in liquid or gas. Ultramicroscopy 111:1650–1658. https://doi.org/https://doi.org/10.1016/j.ultramic.2011.08.001
181. Bogner A, Thollet G, Basset D, et al (2005) Wet STEM: A new development in environmental SEM for imaging nano-objects included in a liquid phase. Ultramicroscopy 104:290–301. https://doi.org/https://doi.org/10.1016/j.ultramic.2005.05.005
182. Bogner A, Jouneau P-H, Thollet G, et al (2007) A history of scanning electron microscopy developments: Towards "wet-STEM" imaging. Micron 38:390–401. https://doi.org/https://doi.org/10.1016/j.micron.2006.06.008
183. Ede JM (2021) Deep learning in electron microscopy. Machine Learning: Science and Technology 2:011004. https://doi.org/10.1088/2632-2153/abd614

184. Horstmann H, Körber C, Sätzler K, et al (2012) Serial section scanning electron microscopy (S3EM) on silicon wafers for ultra-structural volume imaging of cells and tissues. PLOS ONE 7:1–8. https://doi.org/10.1371/journal.pone.0035172
185. PLUK H, STOKES DJ, LICH B, et al (2009) Advantages of indium–tin oxide-coated glass slides in correlative scanning electron microscopy applications of uncoated cultured cells. Journal of Microscopy 233:353–363. https://doi.org/https://doi.org/10.1111/j.1365-2818.2009.03140.x
186. Sawaguchi A, Kamimura T, Yamashita A, et al (2018) Informative three-dimensional survey of cell/tissue architectures in thick paraffin sections by simple low-vacuum scanning electron microscopy. Scientific Reports 8:7479. https://doi.org/10.1038/s41598-018-25840-8
187. Briggman KL, Bock DD (2012) Volume electron microscopy for neuronal circuit reconstruction. Current Opinion in Neurobiology 22:154–161. https://doi.org/https://doi.org/10.1016/j.conb.2011.10.022
188. Kim GH, Gim JW, Lee KJ (2016) Nano-resolution connectomics using large-volume electron microscopy. Applied Microscopy 46:171–175. https://doi.org/10.9729/AM.2016.46.4.171
189. Oho E, Okugawa K, Kawamata S (2000) Practical SEM system based on the montage technique applicable to ultralow-magnification observation, while maintaining original functions. Journal of Electron Microscopy 49:135–141. https://doi.org/10.1093/oxfordjournals.jmicro.a023777
190. Hong W-P, Lee S-W, Choi H-Z (2013) A stitching algorithm for measuring large areas using scanning electron microscopes. International Journal of Precision Engineering and Manufacturing 14:147–151. https://doi.org/10.1007/s12541-013-0020-3
191. Khoonkari N, Anand C, Bassim N (2021) Making the stitching process of montaged SEM images automatic using fourier transform properties. Microscopy and Microanalysis 27:478–480. https://doi.org/10.1017/S1431927621002208
192. Singla A, Lippmann B, Graeb H (2021) Recovery of 2D and 3D layout information through an advanced image stitching algorithm using scanning electron microscope images. In: 2020 25th international conference on pattern recognition (ICPR). pp 3860–3867
193. Titze B, Genoud C, Friedrich RW (2018) SBEMimage: Versatile acquisition control software for serial block-face electron microscopy. Frontiers in Neural Circuits 12:54. https://doi.org/10.3389/fncir.2018.00054
194. Knothe Tate ML, Srikantha A, Wojek C, Zeidler D (2021) Connectomics of bone to brain—probing physical renderings of cellular experience. Frontiers in Physiology 12:1018. https://doi.org/10.3389/fphys.2021.647603
195. EMMENLAUER M, RONNEBERGER O, PONTI A, et al (2009) XuvTools: Free, fast and reliable stitching of large 3D datasets. Journal of Microscopy 233:42–60. https://doi.org/https://doi.org/10.1111/j.1365-2818.2008.03094.x
196. Brantner CA, Rasche M, Burcham KE, et al (2016) A reverse engineering approach for imaging neuronal architecture – large-area, high-resolution SEM imaging. Microscopy Today 24:28–33. https://doi.org/10.1017/S1551929516000730
197. Kataoka M, Ishida K, Ogasawara K, et al (2019) Serial section array scanning electron microscopy analysis of cells from lung autopsy specimens following fatal a/H1N1 2009 pandemic influenza virus infection. Journal of Virology 93:e00644–19. https://doi.org/10.1128/JVI.00644-19
198. More HL, Chen J, Gibson E, et al (2011) A semi-automated method for identifying and measuring myelinated nerve fibers in scanning electron microscope images. Journal of Neuroscience Methods 201:149–158. https://doi.org/https://doi.org/10.1016/j.jneumeth.2011.07.026

199. Lena Eberle A, Schalek R, Lichtman JW, et al (2015) Multiple-beam scanning electron microscopy. Microscopy Today 23:12–19. https://doi.org/10.1017/S1551929515000012
200. EBERLE AL, MIKULA S, SCHALEK R, et al (2015) High-resolution, high-throughput imaging with a multibeam scanning electron microscope. Journal of Microscopy 259:114–120. https://doi.org/https://doi.org/10.1111/jmi.12224
201. Eberle AL, Zeidler D (2018) Multi-beam scanning electron microscopy for high-throughput imaging in connectomics research. Frontiers in Neuroanatomy 12:112. https://doi.org/10.3389/fnana.2018.00112
202. Ren Y, Kruit P (2016) Transmission electron imaging in the delft multibeam scanning electron microscope 1. Journal of Vacuum Science & Technology B 34:06KF02. https://doi.org/10.1116/1.4966216
203. Templier T (2019) MagC, magnetic collection of ultrathin sections for volumetric correlative light and electron microscopy. eLife 8:e45696. https://doi.org/10.7554/eLife.45696
204. Eberle AL, Selchow O, Thaler M, et al (2014) Mission (im)possible – mapping the brain becomes a reality. Microscopy 64:45–55. https://doi.org/10.1093/jmicro/dfu104
205. Shami GJ, Cheng D, Huynh M, et al (2016) 3-d EM exploration of the hepatic microarchitecture – lessons learned from large-volume in situ serial sectioning. Scientific Reports 6:36744. https://doi.org/10.1038/srep36744
206. Ichimura K, Kakuta S, Kawasaki Y, et al (2017) Morphological process of podocyte development revealed by block-face scanning electron microscopy. Journal of Cell Science 130:132–142. https://doi.org/10.1242/jcs.187815
207. Sato Y, Boor P, Fukuma S, et al (2020) Developmental stages of tertiary lymphoid tissue reflect local injury and inflammation in mouse and human kidneys. Kidney International 98:448–463. https://doi.org/10.1016/j.kint.2020.02.023
208. Barajas L (1970) The ultrastructure of the juxtaglomerular apparatus as disclosed by three-dimensional reconstructions from serial sections. The anatomical relationship between the tubular and vascular components. Journal of ultrastructure research 33 1:116–47
209. Barajas L, Müller J (1973) The innervation of the juxtaglomerular apparatus and surrounding tubules: A quantitative analysis by serial section electron microscopy. Journal of Ultrastructure Research 43:107–132. https://doi.org/https://doi.org/10.1016/S0022-5320(73)90073-7
210. DROBNE D, MILANI M, ZRIMEC A, et al (2005) Electron and ion imaging of gland cells using the FIB/SEM system. Journal of Microscopy 219:29–35. https://doi.org/https://doi.org/10.1111/j.1365-2818.2005.01490.x
211. Hirashima S, Kanazawa T, Ohta K, Nakamura K (2020) Three-dimensional ultrastructural imaging and quantitative analysis of the periodontal ligament. Anatomical Science International 95:1–11. https://doi.org/10.1007/s12565-019-00502-5
212. Delgado T, Petralia RS, Freeman DW, et al (2019) Comparing 3D ultrastructure of presynaptic and postsynaptic mitochondria. Biology Open 8: https://doi.org/10.1242/bio.044834
213. YOUNG RJ, DINGLE T, ROBINSON K, PUGH PJA (1993) An application of scanned focused ion beam milling to studies on the internal morphology of small arthropods. Journal of Microscopy 172:81–88. https://doi.org/https://doi.org/10.1111/j.1365-2818.1993.tb03396.x
214. Weigel AV, Chang C-L, Shtengel G, et al (2021) ER-to-golgi protein delivery through an interwoven, tubular network extending from ER. Cell 184:2412–2429.e16. https://doi.org/10.1016/j.cell.2021.03.035

215. Wei D, Jacobs S, Modla S, et al (2012) High-resolution three-dimensional reconstruction of a whole yeast cell using focused-ion beam scanning electron microscopy. BioTechniques 53:41–48. https://doi.org/10.2144/000113850
216. Turegano-Lopez M, Santuy A, DeFelipe J, Merchan-Perez A (2019) Size, Shape, and Distribution of Multivesicular Bodies in the Juvenile Rat Somatosensory Cortex: A 3D Electron Microscopy Study. Cerebral Cortex 30:1887–1901. https://doi.org/10.1093/cercor/bhz211
217. Murata K, Hirata A, Ohta K, et al (2019) Morphometric analysis in mouse scleral fibroblasts using focused ion beam/scanning electron microscopy. Scientific Reports 9:6329. https://doi.org/10.1038/s41598-019-42758-x
218. Robles H, Park S, Joens MS, et al (2019) Characterization of the bone marrow adipocyte niche with three-dimensional electron microscopy. Bone 118:89–98. https://doi.org/https://doi.org/10.1016/j.bone.2018.01.020
219. Hirashima S, Ohta K, Kanazawa T, et al (2016) Three-dimensional ultrastructural analysis of cells in the periodontal ligament using focused ion beam/scanning electron microscope tomography. Scientific Reports 6:39435. https://doi.org/10.1038/srep39435
220. Hirashima S, Ohta K, Kanazawa T, et al (2019) Three-dimensional ultrastructural and histomorphological analysis of the periodontal ligament with occlusal hypofunction via focused ion beam/scanning electron microscope tomography. Scientific Reports 9:9520. https://doi.org/10.1038/s41598-019-45963-w
221. Kawasaki Y, Hosoyamada Y, Miyaki T, et al (2021) Three-dimensional architecture of glomerular endothelial cells revealed by FIB-SEM tomography. Frontiers in Cell and Developmental Biology 9:339. https://doi.org/10.3389/fcell.2021.653472
222. Miyaki T, Kawasaki Y, Hosoyamada Y, et al (2020) Three-dimensional imaging of podocyte ultrastructure using FE-SEM and FIB-SEM tomography. Cell and Tissue Research 379:245–254. https://doi.org/10.1007/s00441-019-03118-3
223. Miyamoto T, Hosoba K, Itabashi T, et al (2020) Insufficiency of ciliary cholesterol in hereditary zellweger syndrome. The EMBO Journal 39:e103499. https://doi.org/https://doi.org/10.15252/embj.2019103499
224. Schneider JP, Wrede C, Mühlfeld C (2020) The three-dimensional ultrastructure of the human alveolar epithelium revealed by focused ion beam electron microscopy. International Journal of Molecular Sciences 21: https://doi.org/10.3390/ijms21031089
225. Ronchi P, Mizzon G, Machado P, et al (2021) High-precision targeting workflow for volume electron microscopy. Journal of Cell Biology 220: https://doi.org/10.1083/jcb.202104069
226. Müller A, Schmidt D, Xu CS, et al (2020) 3D FIB-SEM reconstruction of microtubule–organelle interaction in whole primary mouse β cells. Journal of Cell Biology 220: https://doi.org/10.1083/jcb.202010039
227. Miyazono Y, Hirashima S, Ishihara N, et al (2018) Uncoupled mitochondria quickly shorten along their long axis to form indented spheroids, instead of rings, in a fission-independent manner. Scientific Reports 8:350. https://doi.org/10.1038/s41598-017-18582-6
228. Xu CS, Hayworth KJ, Lu Z, et al (2017) Enhanced FIB-SEM systems for large-volume 3D imaging. eLife 6:e25916. https://doi.org/10.7554/eLife.25916
229. Xu CS, Pang S, Hayworth KJ, Hess HF (2020) Transforming FIB-SEMFocused ion beam scanning electron microscopy (FIB-SEM)systems for large-volume ConnectomicsConnectomics and cell BiologyCell biology. In: Wacker I, Hummel E, Burgold S, Schröder R (eds) Volume microscopy : Multiscale imaging with photons, electrons, and ions. Springer US, New York, NY, pp 221–243

230. Xu CS, Pang S, Shtengel G, et al (2021) An open-access volume electron microscopy atlas of whole cells and tissues. Nature. https://doi.org/10.1038/s41586-021-03992-4
231. Heinrich L, Bennett D, Ackerman D, et al (2021) Whole-cell organelle segmentation in volume electron microscopy. Nature. https://doi.org/10.1038/s41586-021-03977-3
232. KREMER A, LIPPENS S, BARTUNKOVA S, et al (2015) Developing 3D SEM in a broad biological context. Journal of Microscopy 259:80–96. https://doi.org/https://doi.org/10.1111/jmi.12211
233. Denk W, Horstmann H (2004) Serial block-face scanning electron microscopy to reconstruct three-dimensional tissue nanostructure. PLOS Biology 2:null. https://doi.org/10.1371/journal.pbio.0020329
234. Briggman KL, Denk W (2006) Towards neural circuit reconstruction with volume electron microscopy techniques. Current Opinion in Neurobiology 16:562–570. https://doi.org/https://doi.org/10.1016/j.conb.2006.08.010
235. Motta A, Berning M, Boergens KM, et al (2019) Dense connectomic reconstruction in layer 4 of the somatosensory cortex. Science 366: https://doi.org/10.1126/science.aay3134
236. Ichimura K, Miyazaki N, Sadayama S, et al (2015) Three-dimensional architecture of podocytes revealed by block-face scanning electron microscopy. Scientific Reports 5:8993. https://doi.org/10.1038/srep08993
237. Daniel E, Daudé M, Kolotuev I, et al (2018) Coordination of septate junctions assembly and completion of cytokinesis in proliferative epithelial tissues. Current Biology 28:1380–1391.e4. https://doi.org/10.1016/j.cub.2018.03.034
238. Armer HEJ, Mariggi G, Png KMY, et al (2009) Imaging transient blood vessel fusion events in zebrafish by correlative volume electron microscopy. PLOS ONE 4:1–10. https://doi.org/10.1371/journal.pone.0007716
239. Peddie CJ, Collinson LM (2014) Exploring the third dimension: Volume electron microscopy comes of age. Micron 61:9–19. https://doi.org/https://doi.org/10.1016/j.micron.2014.01.009
240. Thaunat O, Granja AG, Barral P, et al (2012) Asymmetric segregation of polarized antigen on b cell division shapes presentation capacity. Science 335:475–479. https://doi.org/10.1126/science.1214100
241. Falkenberg CV, Azeloglu EU, Stothers M, et al (2017) Fragility of foot process morphology in kidney podocytes arises from chaotic spatial propagation of cytoskeletal instability. PLOS Computational Biology 13:1–21. https://doi.org/10.1371/journal.pcbi.1005433
242. Nguyen HB, Thai TQ, Saitoh S, et al (2016) Conductive resins improve charging and resolution of acquired images in electron microscopic volume imaging. Scientific Reports 6:23721. https://doi.org/10.1038/srep23721
243. Randles MJ, Collinson S, Starborg T, et al (2016) Three-dimensional electron microscopy reveals the evolution of glomerular barrier injury. Scientific Reports 6:35068. https://doi.org/10.1038/srep35068
244. Arkill KP, Qvortrup K, Starborg T, et al (2014) Resolution of the three dimensional structure of components of the glomerular filtration barrier. BMC Nephrology 15:24. https://doi.org/10.1186/1471-2369-15-24
245. Miyazaki N, Esaki M, Ogura T, Murata K (2014) Serial block-face scanning electron microscopy for three-dimensional analysis of morphological changes in mitochondria regulated by Cdc48p/p97 ATPase. Journal of Structural Biology 187:187–193. https://doi.org/https://doi.org/10.1016/j.jsb.2014.05.010
246. Murata K, Esaki M, Ogura T, et al (2014) Whole-cell imaging of the budding yeast saccharomyces cerevisiae by high-voltage scanning transmission electron tomography.

Ultramicroscopy 146:39–45. https://doi.org/https://doi.org/10.1016/j.ultramic.2014.05.008
247. Hughes L, Borrett S, Towers K, et al (2017) Patterns of organelle ontogeny through a cell cycle revealed by whole-cell reconstructions using 3D electron microscopy. Journal of Cell Science 130:637–647. https://doi.org/10.1242/jcs.198887
248. Russell MRG, Lerner TR, Burden JJ, et al (2017) 3D correlative light and electron microscopy of cultured cells using serial blockface scanning electron microscopy. Journal of Cell Science 130:278–291. https://doi.org/10.1242/jcs.188433
249. Puhka M, Joensuu M, Vihinen H, et al (2012) Progressive sheet-to-tubule transformation is a general mechanism for endoplasmic reticulum partitioning in dividing mammalian cells. Molecular Biology of the Cell 23:2424–2432. https://doi.org/10.1091/mbc.e10-12-0950
250. Melia CE, Peddie CJ, Jong AWM de, et al (2019) Origins of enterovirus replication organelles established by whole-cell electron microscopy. mBio 10:e00951–19. https://doi.org/10.1128/mBio.00951-19
251. Spiers H, Songhurst H, Nightingale L, et al (2021) Deep learning for automatic segmentation of the nuclear envelope in electron microscopy data, trained with volunteer segmentations. Traffic 22:240–253. https://doi.org/https://doi.org/10.1111/tra.12789
252. Smith SJ (2018) Q&a: Array tomography. BMC Biology 16:98. https://doi.org/10.1186/s12915-018-0560-1
253. Hoffpauir BK, Grimes JL, Mathers PH, Spirou GA (2006) Synaptogenesis of the calyx of held: Rapid onset of function and one-to-one morphological innervation. Journal of Neuroscience 26:5511–5523. https://doi.org/10.1523/JNEUROSCI.5525-05.2006
254. Yamaguchi M, Okada H, Namiki Y (2009) Smart specimen preparation for freeze substitution and serial ultrathin sectioning of yeast cells. Journal of Electron Microscopy 58:261–266. https://doi.org/10.1093/jmicro/dfp013
255. Ylä-Anttila P, Vihinen H, Jokitalo E, Eskelinen E-L (2009) 3D tomography reveals connections between the phagophore and endoplasmic reticulum. Autophagy 5:1180–1185. https://doi.org/10.4161/auto.5.8.10274
256. Zheng Z, Lauritzen JS, Perlman E, et al (2018) A complete electron microscopy volume of the brain of adult <em>drosophila melanogaster</em>. Cell 174:730–743.e22. https://doi.org/10.1016/j.cell.2018.06.019
257. Micheva KD, Smith SJ (2007) Array tomography: A new tool for imaging the molecular architecture and ultrastructure of neural circuits. Neuron 55:25–36. https://doi.org/10.1016/j.neuron.2007.06.014
258. Reichelt M, Joubert L, Perrino J, et al (2012) 3D reconstruction of VZV infected cell nuclei and PML nuclear cages by serial section array scanning electron microscopy and electron tomography. PLOS Pathogens 8:1–17. https://doi.org/10.1371/journal.ppat.1002740
259. Koga D, Kusumi S, Ushiki T (2015) Three-dimensional shape of the Golgi apparatus in different cell types: serial section scanning electron microscopy of the osmium-impregnated Golgi apparatus †. Microscopy 65:145–157. https://doi.org/10.1093/jmicro/dfv360
260. WACKER I, CHOCKLEY P, BARTELS C, et al (2015) Array tomography: Characterizing FAC-sorted populations of zebrafish immune cells by their 3D ultrastructure. Journal of Microscopy 259:105–113. https://doi.org/https://doi.org/10.1111/jmi.12223
261. Kim E, Lee J, Noh S, et al (2020) Double staining method for array tomography using scanning electron microscopy. Applied Microscopy 50:14. https://doi.org/10.1186/s42649-020-00033-8

262. Koike T, Kataoka Y, Maeda M, et al (2017) A device for ribbon collection for array tomography with scanning electron microscopy. ACTA HISTOCHEMICA ET CYTOCHEMICA 50:135–140. https://doi.org/10.1267/ahc.17013
263. Koike T, Yamada H (2019) Methods for array tomography with correlative light and electron microscopy. Medical Molecular Morphology 52:8–14. https://doi.org/10.1007/s00795-018-0194-y
264. Phelps JS, Hildebrand DGC, Graham BJ, et al (2021) Reconstruction of motor control circuits in adult <em>drosophila</em> using automated transmission electron microscopy. Cell 184:759–774.e18. https://doi.org/10.1016/j.cell.2020.12.013
265. Yin W, Brittain D, Borseth J, et al (2020) A petascale automated imaging pipeline for mapping neuronal circuits with high-throughput transmission electron microscopy. Nature Communications 11:4949. https://doi.org/10.1038/s41467-020-18659-3
266. Kislinger G, Gnägi H, Kerschensteiner M, et al (2020) Multiscale ATUM-FIB microscopy enables targeted ultrastructural analysis at isotropic resolution. iScience 23:101290. https://doi.org/https://doi.org/10.1016/j.isci.2020.101290
267. Hayworth K, Kasthuri N, Schalek R, Lichtman J (2006) Automating the collection of ultrathin serial sections for large volume TEM reconstructions. Microscopy and Microanalysis 12:86–87. https://doi.org/10.1017/S1431927606066268
268. Hildebrand DGC, Cicconet M, Torres RM, et al (2017) Whole-brain serial-section electron microscopy in larval zebrafish. Nature 545:345–349. https://doi.org/10.1038/nature22356
269. Morgan JL, Lichtman JW (2020) An individual interneuron participates in many kinds of inhibition and innervates much of the mouse visual thalamus. Neuron 106:468–481.e2. https://doi.org/10.1016/j.neuron.2020.02.001
270. Witvliet D, Mulcahy B, Mitchell JK, et al (2021) Connectomes across development reveal principles of brain maturation. Nature 596:257–261. https://doi.org/10.1038/s41586-021-03778-8
271. Vogelstein JT, Perlman E, Falk B, et al (2018) A community-developed open-source computational ecosystem for big neuro data. Nature Methods 15:846–847. https://doi.org/10.1038/s41592-018-0181-1
272. Hider R, Kleissas DM, Pryor D, et al (2019) The block object storage service (bossDB): A cloud-native approach for petascale neuroscience discovery. bioRxiv. https://doi.org/10.1101/217745
273. Johnson EC, Wilt M, Rodriguez LM, et al (2020) Toward a scalable framework for reproducible processing of volumetric, nanoscale neuroimaging datasets. GigaScience 9: https://doi.org/10.1093/gigascience/giaa147
274. Shapson-Coe A, Januszewski DR Michałand Berger, Pope A, et al (2021) A connectomic study of a petascale fragment of human cerebral cortex. bioRxiv
275. Graham BJ, Hildebrand DGC, Kuan AT, et al (2019) High-throughput transmission electron microscopy with automated serial sectioning. bioRxiv. https://doi.org/10.1101/657346
276. Collman F, Buchanan J, Phend KD, et al (2015) Mapping synapses by conjugate light-electron array tomography. Journal of Neuroscience 35:5792–5807. https://doi.org/10.1523/JNEUROSCI.4274-14.2015
277. Joesch M, Mankus D, Yamagata M, et al (2016) Reconstruction of genetically identified neurons imaged by serial-section electron microscopy. eLife 5:e15015. https://doi.org/10.7554/eLife.15015
278. Boer P de, Hoogenboom JP, Giepmans BNG (2015) Correlated light and electron microscopy: Ultrastructure lights up! Nature Methods 12:503–513. https://doi.org/10.1038/nmeth.3400

279. Oorschot V, Lindsey BW, Kaslin J, Ramm G (2021) TEM, SEM, and STEM-based immuno-CLEM workflows offer complementary advantages. Scientific Reports 11:899. https://doi.org/10.1038/s41598-020-79637-9
280. Luckner M, Burgold S, Filser S, et al (2018) Label-free 3D-CLEM using endogenous tissue landmarks. iScience 6:92–101. https://doi.org/https://doi.org/10.1016/j.isci.2018.07.012
281. Weinhard L, Bartolomei G di, Bolasco G, et al (2018) Microglia remodel synapses by presynaptic trogocytosis and spine head filopodia induction. Nature Communications 9:1228. https://doi.org/10.1038/s41467-018-03566-5
282. Fang T, Lu X, Berger D, et al (2018) Nanobody immunostaining for correlated light and electron microscopy with preservation of ultrastructure. Nature Methods 15:1029–1032. https://doi.org/10.1038/s41592-018-0177-x
283. Booth DG, Beckett AJ, Prior IA, Meijer D (2019) SuperCLEM: an accessible correlative light and electron microscopy approach for investigation of neurons and glia in vitro. Biology Open 8: https://doi.org/10.1242/bio.042085
284. Thomas CI, Ryan MA, Scholl B, et al (2021) Targeting functionally characterized synaptic architecture using inherent fiducials and 3D correlative microscopy. Microscopy and Microanalysis 27:156–169. https://doi.org/10.1017/S1431927620024757
285. Livet J, Weissman TA, Kang H, et al (2007) Transgenic strategies for combinatorial expression of fluorescent proteins in the nervous system. Nature 450:56–62. https://doi.org/10.1038/nature06293
286. Weissman TA, Pan YA (2015) Brainbow: New Resources and Emerging Biological Applications for Multicolor Genetic Labeling and Analysis. Genetics 199:293–306. https://doi.org/10.1534/genetics.114.172510
287. Pan YA, Freundlich T, Weissman TA, et al (2013) Zebrabow: multispectral cell labeling for cell tracing and lineage analysis in zebrafish. Development 140:2835–2846. https://doi.org/10.1242/dev.094631
288. Trzaskoma P, Ruszczycki B, Lee B, et al (2020) Ultrastructural visualization of 3D chromatin folding using volume electron microscopy and DNA in situ hybridization. Nature Communications 11:2120. https://doi.org/10.1038/s41467-020-15987-2
289. Tsang TK, Bushong EA, Boassa D, et al (2018) High-quality ultrastructural preservation using cryofixation for 3D electron microscopy of genetically labeled tissues. eLife 7:e35524. https://doi.org/10.7554/eLife.35524
290. Lam SS, Martell JD, Kamer KJ, et al (2015) Directed evolution of APEX2 for electron microscopy and proximity labeling. Nature Methods 12:51–54. https://doi.org/10.1038/nmeth.3179
291. Martell JD, Deerinck TJ, Lam SS, et al (2017) Electron microscopy using the genetically encoded APEX2 tag in cultured mammalian cells. Nature Protocols 12:1792–1816. https://doi.org/10.1038/nprot.2017.065
292. Ariotti N, Rae J, Giles N, et al (2018) Ultrastructural localisation of protein interactions using conditionally stable nanobodies. PLOS Biology 16:1–11. https://doi.org/10.1371/journal.pbio.2005473
293. Han Y, Branon TC, Martell JD, et al (2019) Directed evolution of split APEX2 peroxidase. ACS Chemical Biology 14:619–635. https://doi.org/10.1021/acschembio.8b00919
294. Mavylutov T, Chen X, Guo L, Yang J (2018) APEX2- tagging of sigma 1-receptor indicates subcellular protein topology with cytosolic n-terminus and ER luminal c-terminus. Protein & Cell 9:733–737. https://doi.org/10.1007/s13238-017-0468-5
295. Goo MS, Sancho L, Slepak N, et al (2017) Activity-dependent trafficking of lysosomes in dendrites and dendritic spines. Journal of Cell Biology 216:2499–2513. https://doi.org/10.1083/jcb.201704068

296. Mavlyutov TA, Yang H, Epstein ML, et al (2017) APEX2-enhanced electron microscopy distinguishes sigma-1 receptor localization in the nucleoplasmic reticulum. Oncotarget 8:51317–51330. https://doi.org/https://doi.org/10.18632/oncotarget.17906
297. Hung V, Lam SS, Udeshi ND, et al (2017) Proteomic mapping of cytosol-facing outer mitochondrial and ER membranes in living human cells by proximity biotinylation. eLife 6:e24463. https://doi.org/10.7554/eLife.24463
298. Lee S-Y, Kang M-G, Park J-S, et al (2016) APEX fingerprinting reveals the subcellular localization of proteins of interest. Cell Reports 15:1837–1847. https://doi.org/10.1016/j.celrep.2016.04.064
299. Adams SR, Mackey MR, Ramachandra R, et al (2016) Multicolor electron microscopy for simultaneous visualization of multiple molecular species. Cell Chemical Biology 23:1417–1427. https://doi.org/10.1016/j.chembiol.2016.10.006
300. Rae J, Ferguson C, Ariotti N, et al (2021) A robust method for particulate detection of a genetic tag for 3D electron microscopy. eLife 10:e64630. https://doi.org/10.7554/eLife.64630
301. Fermie J, Liv N, Brink C ten, et al (2018) Single organelle dynamics linked to 3D structure by correlative live-cell imaging and 3D electron microscopy. Traffic 19:354–369. https://doi.org/https://doi.org/10.1111/tra.12557
302. Betzig E, Patterson GH, Sougrat R, et al (2006) Imaging intracellular fluorescent proteins at nanometer resolution. Science 313:1642–1645. https://doi.org/10.1126/science.1127344
303. Ando T, Bhamidimarri SP, Brending N, et al (2018) The 2018 correlative microscopy techniques roadmap. Journal of Physics D: Applied Physics 51:443001. https://doi.org/10.1088/1361-6463/aad055
304. Kurokawa K, Osakada H, Kojidani T, et al (2019) Visualization of secretory cargo transport within the Golgi apparatus. Journal of Cell Biology 218:1602–1618. https://doi.org/10.1083/jcb.201807194
305. Shimizu Y, Takagi J, Ito E, et al (2021) Cargo sorting zones in the trans-golgi network visualized by super-resolution confocal live imaging microscopy in plants. Nature Communications 12:1901. https://doi.org/10.1038/s41467-021-22267-0
306. Hoffman DP, Shtengel G, Xu CS, et al (2020) Correlative three-dimensional super-resolution and block-face electron microscopy of whole vitreously frozen cells. Science 367:eaaz5357. https://doi.org/10.1126/science.aaz5357
307. Fu Z, Peng D, Zhang M, et al (2020) mEosEM withstands osmium staining and epon embedding for super-resolution CLEM. Nature Methods 17:55–58. https://doi.org/10.1038/s41592-019-0613-6
308. Sakdinawat A, Attwood D (2010) Nanoscale x-ray imaging. Nature Photonics 4:840–848. https://doi.org/10.1038/nphoton.2010.267
309. (1983) Abstracts fifth annual scientific meeting of the american society for bone and mineral research june 5–7, 1993 hotel intercontinental san antonio, texas. Calcified Tissue International 35:631–706. https://doi.org/10.1007/BF02405107
310. Merkle AP, Gelb J (2013) The ascent of 3D x-ray microscopy in the laboratory. Microscopy Today 21:10–15. https://doi.org/10.1017/S1551929513000060
311. Cengiz IF, Oliveira JM, Reis RL (2018) Micro-CT – a digital 3D microstructural voyage into scaffolds: A systematic review of the reported methods and results. Biomaterials Research 22:26. https://doi.org/10.1186/s40824-018-0136-8
312. Arabi H, Kamali Asl AR, Aghamiri SM (2010) The effect of focal spot size on the spatial resolution of variable resolution x-ray CT scanner. Iranian Journal of Radiation Research (Print) 8:37–43

313. Rueckel J, Stockmar M, Pfeiffer F, Herzen J (2014) Spatial resolution characterization of a x-ray microCT system. Applied Radiation and Isotopes 94:230–234. https://doi.org/https://doi.org/10.1016/j.apradiso.2014.08.014
314. Metscher BD (2009) MicroCT for developmental biology: A versatile tool for high-contrast 3D imaging at histological resolutions. Developmental Dynamics 238:632–640. https://doi.org/https://doi.org/10.1002/dvdy.21857
315. Swart P, Wicklein M, Sykes D, et al (2016) A quantitative comparison of micro-CT preparations in dipteran flies. Scientific Reports 6:39380. https://doi.org/10.1038/srep39380
316. Busse M, Müller M, Kimm MA, et al (2018) Three-dimensional virtual histology enabled through cytoplasm-specific x-ray stain for microscopic and nanoscopic computed tomography. Proceedings of the National Academy of Sciences 115:2293–2298. https://doi.org/10.1073/pnas.1720862115
317. Foster T, Falter JL, McCulloch MT, Clode PL (2016) Ocean acidification causes structural deformities in juvenile coral skeletons. Science Advances 2:e1501130. https://doi.org/10.1126/sciadv.1501130
318. Metscher BD (2009) MicroCT for comparative morphology: Simple staining methods allow high-contrast 3D imaging of diverse non-mineralized animal tissues. BMC Physiology 9:11. https://doi.org/10.1186/1472-6793-9-11
319. Castejón D, Alba-Tercedor J, Rotllant G, et al (2018) Micro-computed tomography and histology to explore internal morphology in decapod larvae. Scientific Reports 8:14399. https://doi.org/10.1038/s41598-018-32709-3
320. Metscher B (2013) Biological applications of x-ray microtomography: Imaging microanatomy, molecular expression and organismal diversity. Microsc Anal (Am Ed) 27:13–16
321. Langer M, Peyrin F (2016) 3D x-ray ultra-microscopy of bone tissue. Osteoporosis International 27:441–455. https://doi.org/10.1007/s00198-015-3257-0
322. Chaurand P, Liu W, Borschneck D, et al (2018) Multi-scale x-ray computed tomography to detect and localize metal-based nanomaterials in lung tissues of in vivo exposed mice. Scientific Reports 8:4408. https://doi.org/10.1038/s41598-018-21862-4
323. Masís J, Mankus D, Wolff SBE, et al (2018) A micro-CT-based method for quantitative brain lesion characterization and electrode localization. Scientific Reports 8:5184. https://doi.org/10.1038/s41598-018-23247-z
324. Toyota E, Ogasawara Y, Fujimoto K, et al (2004) Global heterogeneity of glomerular volume distribution in early diabetic nephropathy. Kidney International 66:855–861. https://doi.org/10.1111/j.1523-1755.2004.00816.x
325. Bentley MD, Jorgensen SM, Lerman LO, et al (2007) Visualization of three-dimensional nephron structure with microcomputed tomography. The Anatomical Record 290:277–283. https://doi.org/https://doi.org/10.1002/ar.20422
326. Müller M, Kimm MA, Ferstl S, et al (2018) Nucleus-specific x-ray stain for 3D virtual histology. Scientific Reports 8:17855. https://doi.org/10.1038/s41598-018-36067-y
327. Degenhardt K, Wright AC, Horng D, et al (2010) Rapid 3D phenotyping of cardiovascular development in mouse embryos by micro-CT with iodine staining. Circulation: Cardiovascular Imaging 3:314–322. https://doi.org/10.1161/CIRCIMAGING.109.918482
328. Cole JM, Symes DR, Lopes NC, et al (2018) High-resolution μCT of a mouse embryo using a compact laser-driven x-ray betatron source. Proceedings of the National Academy of Sciences 115:6335–6340. https://doi.org/10.1073/pnas.1802314115
329. Hsu C-W, Wong L, Rasmussen TL, et al (2016) Three-dimensional microCT imaging of mouse development from early post-implantation to early postnatal stages. Developmental Biology 419:229–236. https://doi.org/https://doi.org/10.1016/j.ydbio.2016.09.011

330. Tamura M, Hosoya M, Fujita M, et al (2013) Overdosage of Hand2 causes limb and heart defects in the human chromosomal disorder partial trisomy distal 4q. Human Molecular Genetics 22:2471–2481. https://doi.org/10.1093/hmg/ddt099
331. Dickinson ME, Flenniken AM, Ji X, et al (2016) High-throughput discovery of novel developmental phenotypes. Nature 537:508–514. https://doi.org/10.1038/nature19356
332. Wong MD, Spring S, Henkelman RM (2014) Structural stabilization of tissue for embryo phenotyping using micro-CT with iodine staining. PLOS ONE 8:1–7. https://doi.org/10.1371/journal.pone.0084321
333. Tun WM, Poologasundarampillai G, Bischof H, et al (2021) A massively multi-scale approach to characterizing tissue architecture by synchrotron micro-CT applied to the human placenta. Journal of The Royal Society Interface 18:20210140. https://doi.org/10.1098/rsif.2021.0140
334. Zdora M-C, Thibault P, Kuo W, et al (2020) X-ray phase tomography with near-field speckles for three-dimensional virtual histology. Optica 7:1221–1227. https://doi.org/10.1364/OPTICA.399421
335. Kaneko Y, Shinohara G, Hoshino M, et al (2017) Intact imaging of human heart structure using x-ray phase-contrast tomography. Pediatric Cardiology 38:390–393. https://doi.org/10.1007/s00246-016-1527-z
336. Töpperwien M, Meer F van der, Stadelmann C, Salditt T (2018) Three-dimensional virtual histology of human cerebellum by x-ray phase-contrast tomography. Proceedings of the National Academy of Sciences 115:6940–6945. https://doi.org/10.1073/pnas.1801678115
337. Kunishima N, Takeda Y, Hirose R, et al (2020) Visualization of internal 3D structure of small live seed on germination by laboratory-based x-ray microscopy with phase contrast computed tomography. Plant Methods 16:7. https://doi.org/10.1186/s13007-020-0557-y
338. Handschuh S, Baeumler N, Schwaha T, Ruthensteiner B (2013) A correlative approach for combining microCT, light and transmission electron microscopy in a single 3D scenario. Frontiers in Zoology 10:44. https://doi.org/10.1186/1742-9994-10-44
339. Sengle G, Tufa SF, Sakai LY, et al (2013) A correlative method for imaging identical regions of samples by micro-CT, light microscopy, and electron microscopy: Imaging adipose tissue in a model system. Journal of Histochemistry & Cytochemistry 61:263–271. https://doi.org/10.1369/0022155412473757
340. Parlanti P, Cappello V, Brun F, et al (2017) Size and specimen-dependent strategy for x-ray micro-ct and tem correlative analysis of nervous system samples. Scientific Reports 7:2858. https://doi.org/10.1038/s41598-017-02998-1
341. Burnett TL, McDonald SA, Gholinia A, et al (2014) Correlative tomography. Scientific Reports 4:4711. https://doi.org/10.1038/srep04711
342. Bushong EA, Johnson DD, Kim K-Y, et al (2015) X-ray microscopy as an approach to increasing accuracy and efficiency of serial block-face imaging for correlated light and electron microscopy of biological specimens. Microscopy and Microanalysis 21:231–238. https://doi.org/10.1017/S1431927614013579
343. Mizutani R, Suzuki Y (2012) X-ray microtomography in biology. Micron 43:104–115. https://doi.org/https://doi.org/10.1016/j.micron.2011.10.002
344. Zaqout S, Kaindl AM (2016) Golgi-cox staining step by step. Frontiers in Neuroanatomy 10:38. https://doi.org/10.3389/fnana.2016.00038
345. Karreman MA, Mercier L, Schieber NL, et al (2016) Fast and precise targeting of single tumor cells in vivo by multimodal correlative microscopy. Journal of Cell Science 129:444–456. https://doi.org/10.1242/jcs.181842
346. Müller M, Sena Oliveira I de, Allner S, et al (2017) Myoanatomy of the velvet worm leg revealed by laboratory-based nanofocus x-ray source tomography. Proceedings of the

National Academy of Sciences 114:12378–12383. https://doi.org/10.1073/pnas.1710742114
347. Mizutani R, Saiga R, Takeuchi A, et al (2013) Three-dimensional network of drosophila brain hemisphere. Journal of Structural Biology 184:271–279. https://doi.org/https://doi.org/10.1016/j.jsb.2013.08.012
348. Mizutani R, Saiga R, Ohtsuka M, et al (2016) Three-dimensional x-ray visualization of axonal tracts in mouse brain hemisphere. Scientific Reports 6:35061. https://doi.org/10.1038/srep35061
349. Strotton MC, Bodey AJ, Wanelik K, et al (2018) Optimising complementary soft tissue synchrotron x-ray microtomography for reversibly-stained central nervous system samples. Scientific Reports 8:12017. https://doi.org/10.1038/s41598-018-30520-8
350. Kuan AT, Phelps JS, Thomas LA, et al (2020) Dense neuronal reconstruction through x-ray holographic nano-tomography. Nature Neuroscience 23:1637–1643. https://doi.org/10.1038/s41593-020-0704-9
351. Moscheni C, Malucelli E, Castiglioni S, et al (2019) 3D quantitative and ultrastructural analysis of mitochondria in a model of doxorubicin sensitive and resistant human colon carcinoma cells. Cancers 11: https://doi.org/10.3390/cancers11091254
352. Rawson SD, Maksimcuka J, Withers PJ, Cartmell SH (2020) X-ray computed tomography in life sciences. BMC Biology 18:21. https://doi.org/10.1186/s12915-020-0753-2
353. Hummel E, Guttmann P, Werner S, et al (2013) 3D ultrastructural organization of whole chlamydomonas reinhardtii cells studied by nanoscale soft x-ray tomography. PLOS ONE 7:1–9. https://doi.org/10.1371/journal.pone.0053293
354. Larabell CA, Le Gros MA (2004) X-ray tomography generates 3-d reconstructions of the yeast, saccharomyces cerevisiae, at 60-nm resolution. Molecular Biology of the Cell 15:957–962. https://doi.org/10.1091/mbc.e03-07-0522
355. Weinhardt V, Chen J-H, Ekman AA, et al (2020) Switchable resolution in soft x-ray tomography of single cells. PLOS ONE 15:1–14. https://doi.org/10.1371/journal.pone.0227601
356. Le Gros MA, McDermott G, Cinquin BP, et al (2014) Biological soft X-ray tomography on beamline 2.1 at the Advanced Light Source. Journal of Synchrotron Radiation 21:1370–1377. https://doi.org/10.1107/S1600577514015033
357. Dyer EL, Gray Roncal W, Prasad JA, et al (2017) Quantifying mesoscale neuroanatomy using x-ray microtomography. eNeuro 4: https://doi.org/10.1523/ENEURO.0195-17.2017
358. Lin Y-C, Hwu Y, Huang G-S, et al (2017) Differential synchrotron x-ray imaging markers based on the renal microvasculature for tubulointerstitial lesions and glomerulopathy. Scientific Reports 7:3488. https://doi.org/10.1038/s41598-017-03677-x
359. Nango N, Kubota S, Hasegawa T, et al (2016) Osteocyte-directed bone demineralization along canaliculi. Bone 84:279–288. https://doi.org/https://doi.org/10.1016/j.bone.2015.12.006
360. Andrews JC, Almeida E, Meulen MCH van der, et al (2010) Nanoscale x-ray microscopic imaging of mammalian mineralized tissue. Microscopy and Microanalysis 16:327–336. https://doi.org/10.1017/S1431927610000231
361. Pishak VP, Tymochko KB, Antoniuk OP (2002) Automative morphome analysis of medical-biological images. In: Angelsky OV (ed) Selected papers from fifth international conference on correlation optics. International Society for Optics; Photonics; SPIE, pp 411–413
362. Lucocq JM, Mayhew TM, Schwab Y, et al (2015) Systems biology in 3D space – enter the morphome. Trends in Cell Biology 25:59–64. https://doi.org/https://doi.org/10.1016/j.tcb.2014.09.008

363. Mayhew TM (2015) Morphomics: An integral part of systems biology of the human placenta. Placenta 36:329–340. https://doi.org/https://doi.org/10.1016/j.placenta.2015.01.001
364. Kume S, Murakawa Y (2020) Large-area imaging technology of tissue sections using SEM and prospects for comprehensive morphological analysis of biological tissues. KENBIKYO (Japanese) 55:13–17. https://doi.org/10.11410/kenbikyo.55.1_13
365. Heuser J (2000) How to convert a traditional electron microscopy laboratory to digital imaging: Follow the “middle road.” Traffic 1:614–621. https://doi.org/https://doi.org/10.1034/j.1600-0854.2000.010805.x
366. Eliceiri KW, Berthold MR, Goldberg IG, et al (2012) Biological imaging software tools. Nature Methods 9:697–710. https://doi.org/10.1038/nmeth.2084
367. Swedlow JR, Kankaanpää P, Sarkans U, et al (2021) A global view of standards for open image data formats and repositories. Nature Methods. https://doi.org/10.1038/s41592-021-01113-7
368. Swedlow JR, Goldberg IG, Eliceiri KW and (2009) Bioimage informatics for experimental biology. Annual Review of Biophysics 38:327–346. https://doi.org/10.1146/annurev.biophys.050708.133641
369. Kobayashi N, Kume S, Lenz K, Masuya H (2018) RIKEN MetaDatabase: A database platform for health care and life sciences as a microcosm of linked open data cloud. Int J Semantic Web Inf Syst 14:140–164
370. Kamdar MR, Musen MA (2021) An empirical meta-analysis of the life sciences linked open data on the web. Scientific Data 8:24. https://doi.org/10.1038/s41597-021-00797-y
371. Ellenberg J, Swedlow JR, Barlow M, et al (2018) A call for public archives for biological image data. Nature Methods 15:849–854. https://doi.org/10.1038/s41592-018-0195-8
372. Swedlow JR (2007) The open microscopy environment: A collaborative data modeling and software development project for biological image informatics. In: Shorte SL, Frischknecht F (eds) Imaging cellular and molecular biological functions. Springer Berlin Heidelberg, Berlin, Heidelberg, pp 71–92
373. Williams E, Moore J, Li SW, et al (2017) Image data resource: A bioimage data integration and publication platform. Nature Methods 14:775–781. https://doi.org/10.1038/nmeth.4326
374. Swedlow J (2020) Open microscopy environment: OME is a consortium of universities, research labs, industry and developers producing open-source software and format standards for microscopy data.
375. Allan C, Burel J-M, Moore J, et al (2012) OMERO: Flexible, model-driven data management for experimental biology. Nature Methods 9:245–253. https://doi.org/10.1038/nmeth.1896
376. Goldberg IG, Allan C, Burel J-M, et al (2005) The open microscopy environment (OME) data model and XML file: Open tools for informatics and quantitative analysis in biological imaging. Genome Biology 6:R47. https://doi.org/10.1186/gb-2005-6-5-r47
377. Moore J, Allan C, Besson S, et al (2021) OME-NGFF: Scalable format strategies for interoperable bioimaging data. bioRxiv
378. Little S, Hunter J (2004) Rules-by-example – a novel approach to semantic indexing and querying of images. In: McIlraith SA, Plexousakis D, Harmelen F van (eds) The semantic web – ISWC 2004. Springer Berlin Heidelberg, Berlin, Heidelberg, pp 534–548
379. Kobayashi N, Moore J, Onami S, Swedlow J (2019) OME core ontology: An OWL-based life science imaging data model. In: SWAT4HCLS
380. Kume S, Masuya H, Kataoka Y, Kobayashi N (2016) Development of an ontology for an integrated image analysis platform to enable global sharing of microscopy imaging data. In: International semantic web conference

381. Sarkans U, Chiu W, Collinson L, et al (2021) REMBI: Recommended metadata for biological images—enabling reuse of microscopy data in biology. Nature Methods. https://doi.org/10.1038/s41592-021-01166-8
382. Kobayashi N, Kume S, Moore J, Swedlow JR (2018) OME Ontology: A Novel Data and Tool Integration Methodology for Multi-Modal Imaging in the Life Sciences. Proceeding of SWAT4HCLS. https://doi.org/10.6084/m9.figshare.7325063.v1
383. Walter A, Paul-Gilloteaux P, Plochberger B, et al (2020) Correlated multimodal imaging in life sciences: Expanding the biomedical horizon. Frontiers in Physics 8:47. https://doi.org/10.3389/fphy.2020.00047
384. Nelson G, Boehm U, Bagley S, et al QUAREP-LiMi: A community-driven initiative to establish guidelines for quality assessment and reproducibility for instruments and images in light microscopy. Journal of Microscopy n/a: https://doi.org/https://doi.org/10.1111/jmi.13041
385. Miron E, Oldenkamp R, Brown JM, et al (2020) Chromatin arranges in chains of mesoscale domains with nanoscale functional topography independent of cohesin. Science Advances 6: https://doi.org/10.1126/sciadv.aba8811
386. King GA, Goodman JS, Schick JG, et al (2019) Meiotic cellular rejuvenation is coupled to nuclear remodeling in budding yeast. eLife 8:e47156. https://doi.org/10.7554/eLife.47156
387. Tagari M, Newman R, Chagoyen M, et al (2002) New electron microscopy database and deposition system. Trends in Biochemical Sciences 27:589. https://doi.org/https://doi.org/10.1016/S0968-0004(02)02176-X
388. Lawson CL, Patwardhan A, Baker ML, et al (2015) EMDataBank unified data resource for 3DEM. Nucleic Acids Research 44:D396–D403. https://doi.org/10.1093/nar/gkv1126
389. Iudin A, Korir PK, Salavert-Torres J, et al (2016) EMPIAR: A public archive for raw electron microscopy image data. Nature Methods 13:387–388. https://doi.org/10.1038/nmeth.3806
390. Karabağ C, Jones ML, Reyes-Aldasoro CC (2021) Volumetric semantic instance segmentation of the plasma membrane of HeLa cells. Journal of Imaging 7: https://doi.org/10.3390/jimaging7060093
391. KOGA D, KUSUMI S, USHIKI T, WATANABE T (2017) <B>integrative method for three-dimensional imaging of the entire golgi apparatus by combining thiamine pyrophosphatase cytochemistry and array tomography using backscattered electron-mode scanning electron micr</b><b>oscopy </b>. Biomedical Research 38:285–296. https://doi.org/10.2220/biomedres.38.285
392. Rogowska J (2009) Chapter 5 - overview and fundamentals of medical image segmentation. In: BANKMAN IN (ed) Handbook of medical image processing and analysis (second edition), Second Edition. Academic Press, Burlington, pp 73–90
393. Liu W, Shi H, He X, et al (2019) An application of optimized otsu multi-threshold segmentation based on fireworks algorithm in cement SEM image. Journal of Algorithms & Computational Technology 13:1748301818797025. https://doi.org/10.1177/1748301818797025
394. Win KY, Choomchuay S, Hamamoto K, Raveesunthornkiat M (2018) Comparative study on automated cell nuclei segmentation methods for cytology pleural effusion images. Journal of Healthcare Engineering 2018:9240389. https://doi.org/10.1155/2018/9240389
395. Min S, Lee B, Yoon S (2016) Deep learning in bioinformatics. Briefings in Bioinformatics 18:851–869. https://doi.org/10.1093/bib/bbw068
396. Chamier L von, Laine RF, Jukkala J, et al (2021) Democratising deep learning for microscopy with ZeroCostDL4Mic. Nature Communications 12:2276. https://doi.org/10.1038/s41467-021-22518-0

397. Laak J van der, Litjens G, Ciompi F (2021) Deep learning in histopathology: The path to the clinic. Nature Medicine 27:775–784. https://doi.org/10.1038/s41591-021-01343-4
398. Barisoni L, Lafata KJ, Hewitt SM, et al (2020) Digital pathology and computational image analysis in nephropathology. Nature Reviews Nephrology 16:669–685. https://doi.org/10.1038/s41581-020-0321-6
399. Srinidhi CL, Ciga O, Martel AL (2021) Deep neural network models for computational histopathology: A survey. Medical Image Analysis 67:101813. https://doi.org/https://doi.org/10.1016/j.media.2020.101813
400. Meijering E (2020) A bird's-eye view of deep learning in bioimage analysis. Computational and Structural Biotechnology Journal 18:2312–2325. https://doi.org/https://doi.org/10.1016/j.csbj.2020.08.003
401. Litjens G, Kooi T, Bejnordi BE, et al (2017) A survey on deep learning in medical image analysis. Medical Image Analysis 42:60–88. https://doi.org/https://doi.org/10.1016/j.media.2017.07.005
402. Wang S, Yang DM, Rong R, et al (2019) Pathology image analysis using segmentation deep learning algorithms. The American Journal of Pathology 189:1686–1698. https://doi.org/https://doi.org/10.1016/j.ajpath.2019.05.007
403. Kaynig V, Fuchs T, Buhmann J (2010) Neuron geometry extraction by perceptual grouping in ssTEM images. In: Computer Vision and Pattern Recognition (CVPR), 2010 IEEE Conference on. IEEE. pp 2902–2909
404. Arganda-Carreras I, Kaynig V, Rueden C, et al (2017) Trainable Weka Segmentation: a machine learning tool for microscopy pixel classification. Bioinformatics 33:2424–2426. https://doi.org/10.1093/bioinformatics/btx180
405. Schindelin J, Rueden CT, Hiner MC, Eliceiri KW (2015) The ImageJ ecosystem: An open platform for biomedical image analysis. Molecular Reproduction and Development 82:518–529. https://doi.org/https://doi.org/10.1002/mrd.22489
406. Turaga SC, Briggman KL, Helmstaedter M, et al (2009) Maximin affinity learning of image segmentation. In: Advances in neural information processing systems 22 - proceedings of the 2009 conference. pp 1865–1873
407. Turaga SC, Murray JF, Jain V, et al (2010) Convolutional Networks Can Learn to Generate Affinity Graphs for Image Segmentation. Neural Computation 22:511–538. https://doi.org/10.1162/neco.2009.10-08-881
408. Cireşan DC, Giusti A, Gambardella LM, Schmidhuber J (2013) Mitosis detection in breast cancer histology images with deep neural networks. In: Mori K, Sakuma I, Sato Y, et al (eds) Medical image computing and computer-assisted intervention – MICCAI 2013. Springer Berlin Heidelberg, Berlin, Heidelberg, pp 411–418
409. Sadanandan SK, Ranefall P, Le Guyader S, Wählby C (2017) Automated training of deep convolutional neural networks for cell segmentation. Scientific Reports 7:7860. https://doi.org/10.1038/s41598-017-07599-6
410. Serag A, Ion-Margineanu A, Qureshi H, et al (2019) Translational AI and deep learning in diagnostic pathology. Frontiers in Medicine 6:185. https://doi.org/10.3389/fmed.2019.00185
411. Lin L, Wang W, Chen B (2019) Leukocyte recognition with convolutional neural network. Journal of Algorithms & Computational Technology 13:1748301818813322. https://doi.org/10.1177/1748301818813322
412. Banik PP, Saha R, Kim K-D (2020) An automatic nucleus segmentation and CNN model based classification method of white blood cell. Expert Systems with Applications 149:113211. https://doi.org/https://doi.org/10.1016/j.eswa.2020.113211
413. Ronneberger O, Fischer P, Brox T (2015) U-net: Convolutional networks for biomedical image segmentation. In: Navab N, Hornegger J, Wells WM, Frangi AF (eds) Medical image computing and computer-assisted intervention – MICCAI 2015. Springer International Publishing, Cham, pp 234–241
414. Falk T, Mai D, Bensch R, et al (2019) U-net: Deep learning for cell counting, detection, and morphometry. Nature Methods 16:67–70. https://doi.org/10.1038/s41592-018-0261-2
415. Quan TM, Hildebrand DGC, Jeong W-K (2021) FusionNet: A deep fully residual convolutional neural network for image segmentation in connectomics. Frontiers in Computer Science 3:34. https://doi.org/10.3389/fcomp.2021.613981
416. Long F (2020) Microscopy cell nuclei segmentation with enhanced u-net. BMC Bioinformatics 21:8. https://doi.org/10.1186/s12859-019-3332-1
417. Wagner FH, Dalagnol R, Tarabalka Y, et al (2020) U-net-id, an instance segmentation model for building extraction from satellite images—case study in the joanópolis city, brazil. Remote Sensing 12: https://doi.org/10.3390/rs12101544
418. Zhao P, Zhang J, Fang W, Deng S (2020) SCAU-net: Spatial-channel attention u-net for gland segmentation. Frontiers in Bioengineering and Biotechnology 8:670. https://doi.org/10.3389/fbioe.2020.00670
419. Zhang K, Zhang H, Zhou H, et al (2019) Zebrafish embryo vessel segmentation using a novel dual ResUNet model. Computational Intelligence and Neuroscience 2019:8214975. https://doi.org/10.1155/2019/8214975
420. Çiçek Ö, Abdulkadir A, Lienkamp SS, et al (2016) 3D u-net: Learning dense volumetric segmentation from sparse annotation. In: Ourselin S, Joskowicz L, Sabuncu MR, et al (eds) Medical image computing and computer-assisted intervention – MICCAI 2016. Springer International Publishing, Cham, pp 424–432
421. Goubran M, Ntiri EE, Akhavein H, et al (2020) Hippocampal segmentation for brains with extensive atrophy using three-dimensional convolutional neural networks. Human Brain Mapping 41:291–308. https://doi.org/https://doi.org/10.1002/hbm.24811
422. Zhang D, Banalagay R, Wang J, et al (2019) Two-level training of a 3D U-Net for accurate segmentation of the intra-cochlear anatomy in head CTs with limited ground truth training data. In: Angelini ED, Landman BA (eds) Medical imaging 2019: Image processing. International Society for Optics; Photonics; SPIE, pp 45–52
423. Siddique N, Paheding S, Elkin CP, Devabhaktuni V (2021) U-net and its variants for medical image segmentation: A review of theory and applications. IEEE Access 9:82031–82057. https://doi.org/10.1109/ACCESS.2021.3086020
424. Bankhead P, Loughrey MB, Fernández JA, et al (2017) QuPath: Open source software for digital pathology image analysis. Scientific Reports 7:16878. https://doi.org/10.1038/s41598-017-17204-5
425. Belevich I, Joensuu M, Kumar D, et al (2016) Microscopy image browser: A platform for segmentation and analysis of multidimensional datasets. PLOS Biology 14:1–13. https://doi.org/10.1371/journal.pbio.1002340
426. Yang L, Ghosh RP, Franklin JM, et al (2020) NuSeT: A deep learning tool for reliably separating and analyzing crowded cells. PLOS Computational Biology 16:1–20. https://doi.org/10.1371/journal.pcbi.1008193
427. Urakubo H, Bullmann T, Kubota Y, et al (2019) UNI-EM: An environment for deep neural network-based automated segmentation of neuronal electron microscopic images. Scientific Reports 9:19413. https://doi.org/10.1038/s41598-019-55431-0
428. Belevich I, Jokitalo E (2021) DeepMIB: User-friendly and open-source software for training of deep learning network for biological image segmentation. PLOS Computational Biology 17:1–9. https://doi.org/10.1371/journal.pcbi.1008374
429. Haberl MG, Churas C, Tindall L, et al (2018) CDeep3M—plug-and-play cloud-based deep learning for image segmentation. Nature Methods 15:677–680. https://doi.org/10.1038/s41592-018-0106-z

430. Stringer C, Wang T, Michaelos M, Pachitariu M (2021) Cellpose: A generalist algorithm for cellular segmentation. Nature Methods 18:100–106. https://doi.org/10.1038/s41592-020-01018-x
431. Tustison NJ, Cook PA, Holbrook AJ, et al (2021) The ANTsX ecosystem for quantitative biological and medical imaging. Scientific Reports 11:9068. https://doi.org/10.1038/s41598-021-87564-6
432. Kume S, Nishida K (2021) BioImageDbs: Bio- and biomedical imaging dataset for machine learning and deep learning (for ExperimentHub)
433. Wei D, Lin Z, Franco-Barranco D, et al (2020) MitoEM dataset: Large-scale 3D mitochondria instance segmentation from EM images. In: Martel AL, Abolmaesumi P, Stoyanov D, et al (eds) Medical image computing and computer assisted intervention – MICCAI 2020. Springer International Publishing, Cham, pp 66–76
434. Khadangi A, Boudier T, Rajagopal V (2021) EM-net: Deep learning for electron microscopy image segmentation. In: 2020 25th international conference on pattern recognition (ICPR). pp 31–38
435. Zeng T, Wu B, Ji S (2017) DeepEM3D: approaching human-level performance on 3D anisotropic EM image segmentation. Bioinformatics 33:2555–2562. https://doi.org/10.1093/bioinformatics/btx188
436. Lee K, Zung J, Li P, et al (2017) Superhuman accuracy on the SNEMI3D connectomics challenge. arXiv preprint arXiv:170600120
437. Januszewski M, Kornfeld J, Li PH, et al (2018) High-precision automated reconstruction of neurons with flood-filling networks. Nature Methods 15:605–610. https://doi.org/10.1038/s41592-018-0049-4
438. Sheridan A, Nguyen T, Deb D, et al (2021) Local shape descriptors for neuron segmentation. bioRxiv. https://doi.org/10.1101/2021.01.18.427039
439. Karabağ C, Jones ML, Peddie CJ, et al (2020) Semantic segmentation of HeLa cells: An objective comparison between one traditional algorithm and four deep-learning architectures. PLOS ONE 15:1–21. https://doi.org/10.1371/journal.pone.0230605
440. Konishi K, Mimura M, Nonaka T, et al (2019) Practical method of cell segmentation in electron microscope image stack using deep convolutional neural network☆. Microscopy 68:338–341. https://doi.org/10.1093/jmicro/dfz016
441. Roels J, De Vylder J, Aelterman J, et al (2017) Convolutional neural network pruning to accelerate membrane segmentation in electron microscopy. In: 2017 IEEE 14th international symposium on biomedical imaging (ISBI 2017). pp 633–637
442. Xiao C, Chen X, Li W, et al (2018) Automatic mitochondria segmentation for EM data using a 3D supervised convolutional network. Frontiers in Neuroanatomy 12:92. https://doi.org/10.3389/fnana.2018.00092
443. Oztel I, Yolcu G, Ersoy I, et al (2017) Mitochondria segmentation in electron microscopy volumes using deep convolutional neural network. In: 2017 IEEE international conference on bioinformatics and biomedicine (BIBM). pp 1195–1200
444. Konishi K, Nonaka T, Takei S, et al (2021) Reducing manual operation time to obtain a segmentation learning model for volume electron microscopy using stepwise deep learning with manual correction. Microscopy. https://doi.org/10.1093/jmicro/dfab025
445. Conrad R, Narayan K (2021) CEM500K, a large-scale heterogeneous unlabeled cellular electron microscopy image dataset for deep learning. eLife 10:e65894. https://doi.org/10.7554/eLife.65894
446. Jiang Y, Li L, Chen X, et al (2021) Three-dimensional ATUM-SEM reconstruction and analysis of hepatic endoplasmic reticulum–organelle interactions. Journal of Molecular Cell Biology. https://doi.org/10.1093/jmcb/mjab032

447. Datta A, Ng KF, Balakrishnan D, et al (2021) A data reduction and compression description for high throughput time-resolved electron microscopy. Nature Communications 12:664. https://doi.org/10.1038/s41467-020-20694-z
448. Zhu J-Y, Park T, Isola P, Efros AA (2017) Unpaired image-to-image translation using cycle-consistent adversarial networks. In: 2017 IEEE international conference on computer vision (ICCV). pp 2242–2251
449. Creswell A, White T, Dumoulin V, et al (2018) Generative adversarial networks: An overview. IEEE Signal Processing Magazine 35:53–65. https://doi.org/10.1109/MSP.2017.2765202
450. Wang F, Henninen TR, Keller D, Erni R (2020) Noise2Atom: Unsupervised denoising for scanning transmission electron microscopy images. Applied Microscopy 50:23. https://doi.org/10.1186/s42649-020-00041-8
451. Mohan S, Manzorro R, Vincent JL, et al (2020) Deep denoising for scientific discovery: A case study in electron microscopy. ArXiv abs/2010.12970:
452. Minnen D, Januszewski M, Shapson-Coe A, et al (2021) Denoising-based image compression for connectomics. bioRxiv. https://doi.org/10.1101/2021.05.29.445828
453. Fang L, Monroe F, Novak SW, et al (2021) Deep learning-based point-scanning super-resolution imaging. Nature Methods 18:406–416. https://doi.org/10.1038/s41592-021-01080-z
454. Haan K de, Ballard ZS, Rivenson Y, et al (2019) Resolution enhancement in scanning electron microscopy using deep learning. Scientific Reports 9:12050. https://doi.org/10.1038/s41598-019-48444-2
455. Januszewski M, Jain V (2019) Segmentation-enhanced CycleGAN. bioRxiv. https://doi.org/10.1101/548081
456. Hagita K, Higuchi T, Jinnai H (2018) Super-resolution for asymmetric resolution of FIB-SEM 3D imaging using AI with deep learning. Scientific Reports 8:5877. https://doi.org/10.1038/s41598-018-24330-1
457. Mi L, Wang H, Meirovitch Y, et al (2020) Learning guided electron microscopy with active acquisition. In: Martel AL, Abolmaesumi P, Stoyanov D, et al (eds) Medical image computing and computer assisted intervention – MICCAI 2020. Springer International Publishing, Cham, pp 77–87
458. Xu Z, Li X, Zhu X, et al (2020) Effective immunohistochemistry pathology microscopy image generation using CycleGAN. Frontiers in Molecular Biosciences 7:243. https://doi.org/10.3389/fmolb.2020.571180
459. Zhang Y, Haan K de, Rivenson Y, et al (2020) Digital synthesis of histological stains using micro-structured and multiplexed virtual staining of label-free tissue. Light: Science & Applications 9:78. https://doi.org/10.1038/s41377-020-0315-y
460. Tsuda H, Hotta K (2019) Cell image segmentation by integrating Pix2pixs for each class. In: 2019 IEEE/CVF conference on computer vision and pattern recognition workshops (CVPRW). pp 1065–1073
461. Rana A, Lowe A, Lithgow M, et al (2020) Use of Deep Learning to Develop and Analyze Computational Hematoxylin and Eosin Staining of Prostate Core Biopsy Images for Tumor Diagnosis. JAMA Network Open 3:e205111–e205111. https://doi.org/10.1001/jamanetworkopen.2020.5111
462. Haan K de, Zhang Y, Zuckerman JE, et al (2021) Deep learning-based transformation of h&e stained tissues into special stains. Nature Communications 12:4884. https://doi.org/10.1038/s41467-021-25221-2
463. Shorten C, Khoshgoftaar TM (2019) A survey on image data augmentation for deep learning. Journal of Big Data 6:60. https://doi.org/10.1186/s40537-019-0197-0

464. Wang D, Khosla A, Gargeya R, et al (2016) Deep learning for identifying metastatic breast cancer. ArXiv abs/1606.05718:
465. Ehteshami Bejnordi B, Veta M, Johannes van Diest P, et al (2017) Diagnostic Assessment of Deep Learning Algorithms for Detection of Lymph Node Metastases in Women With Breast Cancer. JAMA 318:2199–2210. https://doi.org/10.1001/jama.2017.14585
466. Campanella G, Hanna MG, Geneslaw L, et al (2019) Clinical-grade computational pathology using weakly supervised deep learning on whole slide images. Nature Medicine 25:1301–1309. https://doi.org/10.1038/s41591-019-0508-1
467. Tabibu S, Vinod PK, Jawahar CV (2019) Pan-renal cell carcinoma classification and survival prediction from histopathology images using deep learning. Scientific Reports 9:10509. https://doi.org/10.1038/s41598-019-46718-3
468. Coudray N, Ocampo PS, Sakellaropoulos T, et al (2018) Classification and mutation prediction from non–small cell lung cancer histopathology images using deep learning. Nature Medicine 24:1559–1567. https://doi.org/10.1038/s41591-018-0177-5
469. Esteva A, Kuprel B, Novoa RA, et al (2017) Dermatologist-level classification of skin cancer with deep neural networks. Nature 542:115–118. https://doi.org/10.1038/nature21056
470. Kather JN, Pearson AT, Halama N, et al (2019) Deep learning can predict microsatellite instability directly from histology in gastrointestinal cancer. Nature Medicine 25:1054–1056. https://doi.org/10.1038/s41591-019-0462-y
471. Bera K, Schalper KA, Rimm DL, et al (2019) Artificial intelligence in digital pathology — new tools for diagnosis and precision oncology. Nature Reviews Clinical Oncology 16:703–715. https://doi.org/10.1038/s41571-019-0252-y
472. Kolachalama VB, Singh P, Lin CQ, et al (2018) Association of pathological fibrosis with renal survival using deep neural networks. Kidney International Reports 3:464–475. https://doi.org/https://doi.org/10.1016/j.ekir.2017.11.002
473. Hermsen M, Bel T de, Boer M den, et al (2019) Deep learningbased histopathologic assessment of kidney tissue. Journal of the American Society of Nephrology 30:1968–1979. https://doi.org/10.1681/ASN.2019020144
474. Bouteldja N, Klinkhammer BM, Bülow RD, et al (2021) Deep learningbased segmentation and quantification in experimental kidney histopathology. Journal of the American Society of Nephrology 32:52–68. https://doi.org/10.1681/ASN.2020050597
475. Jayapandian CP, Chen Y, Janowczyk AR, et al (2021) Development and evaluation of deep learning–based segmentation of histologic structures in the kidney cortex with multiple histologic stains. Kidney International 99:86–101. https://doi.org/10.1016/j.kint.2020.07.044
476. Tavolara TE, Niazi MKK, Gower AC, et al (2021) Deep learning predicts gene expression as an intermediate data modality to identify susceptibility patterns in <em>mycobacterium tuberculosis</em> infected diversity outbred mice. EBioMedicine 67: https://doi.org/10.1016/j.ebiom.2021.103388
477. Levy-Jurgenson A, Tekpli X, Kristensen VN, Yakhini Z (2020) Spatial transcriptomics inferred from pathology whole-slide images links tumor heterogeneity to survival in breast and lung cancer. Scientific Reports 10:18802. https://doi.org/10.1038/s41598-020-75708-z
478. Echle A, Rindtorff NT, Brinker TJ, et al (2021) Deep learning in cancer pathology: A new generation of clinical biomarkers. British Journal of Cancer 124:686–696. https://doi.org/10.1038/s41416-020-01122-x
479. Murchan P, Ó'Brien C, O'Connell S, et al (2021) Deep learning of histopathological features for the prediction of tumour molecular genetics. Diagnostics 11: https://doi.org/10.3390/diagnostics11081406

480. Wang X, Zou C, Zhang Y, et al (2021) Prediction of BRCA gene mutation in breast cancer based on deep learning and histopathology images. Frontiers in Genetics 12:1147. https://doi.org/10.3389/fgene.2021.661109
481. Brück OE, Lallukka-Brück SE, Hohtari HR, et al (2021) Machine learning of bone marrow histopathology identifies genetic and clinical determinants in patients with MDS. Blood Cancer Discovery 2:238–249. https://doi.org/10.1158/2643-3230.BCD-20-0162
482. Ash JT, Darnell G, Munro D, Engelhardt BE (2021) Joint analysis of expression levels and histological images identifies genes associated with tissue morphology. Nature Communications 12:1609. https://doi.org/10.1038/s41467-021-21727-x
483. Schmauch B, Romagnoni A, Pronier E, et al (2020) A deep learning model to predict RNA-seq expression of tumours from whole slide images. Nature Communications 11:3877. https://doi.org/10.1038/s41467-020-17678-4
484. Howard FM, Dolezal J, Kochanny S, et al (2021) The impact of site-specific digital histology signatures on deep learning model accuracy and bias. Nature Communications 12:4423. https://doi.org/10.1038/s41467-021-24698-1
485. Diao JA, Wang JK, Chui WF, et al (2021) Human-interpretable image features derived from densely mapped cancer pathology slides predict diverse molecular phenotypes. Nature Communications 12:1613. https://doi.org/10.1038/s41467-021-21896-9
486. Helmstaedter M, Briggman KL, Turaga SC, et al (2013) Connectomic reconstruction of the inner plexiform layer in the mouse retina. Nature 500:168–174. https://doi.org/10.1038/nature12346
487. Greener JG, Kandathil SM, Moffat L, Jones DT (2021) A guide to machine learning for biologists. Nature Reviews Molecular Cell Biology. https://doi.org/10.1038/s41580-021-00407-0
488. Yamamoto Y, Tsuzuki T, Akatsuka J, et al (2019) Automated acquisition of explainable knowledge from unannotated histopathology images. Nature Communications 10:5642. https://doi.org/10.1038/s41467-019-13647-8
489. Hecht H, Sarhan MH, Popovici V (2020) Disentangled autoencoder for cross-stain feature extraction in pathology image analysis. Applied Sciences 10: https://doi.org/10.3390/app10186427
490. Stepec D, Skocaj D (2021) Unsupervised detection of cancerous regions in histology imagery using image-to-image translation. In: Proceedings of the IEEE/CVF conference on computer vision and pattern recognition (CVPR) workshops. pp 3785–3792

Supplementary Table 1. Representative list of current public electron microscopy datasets

| Data repository | Dataset title | Biosamples | Dataset ID | Imaging method | Dataset size (format) | DOI or URL |
|---|---|---|---|---|---|---|
| CIL | Perturbation of the glomerular endothelial surface layer results in albumin filtration | Kidney section (100 nm) of adult *Mus musculus* | CCDB:9523 | Mosaic TEM (81x76 tiles) | 17.7GB (TIFF) | https://doi.org/doi:10.7295/W9CCDB9523 |
| | | | CCDB:9525 | Mosaic TEM (69x95 tiles) | 19.1GB (TIFF) | https://doi.org/doi:10.7295/W9CCDB9525 |
| CIL | Serial section reconstruction of functionally characterized neurons | Visual cortex section (45 nm) of adult *Mus musculus* brain | CCDB:8448 | Serial section TEM (1,153 sections) | 271GB (tar) | https://doi.org/doi:10.7295/W9CCDB8448 |
| CIL | Ultrastructural Characterization of the Mouse Optic nerve Head and Retina | Retina section of adult *Mus musculus* | CCDB:7742 | SBF-SEM (206 sections) | 22.5GB (tar) | https://doi.org/doi:10.7295/W9CCDB7742 |
| CIL | 3D reconstruction using serial section scanning electron microscopy of the optic nerve head in mouse | Optic nerve head section of adult *Mus musculus* | CCDB:8391 | SBF-SEM (724 sections) | 136GB (rec) | https://doi.org/doi:10.7295/W9CCDB8391 |
| IDR | Virtual nanoscopy of the mouse glomerulus, mouse embryonic fibroblasts, human dendritic cells and a zebrafish embryo slice | Tissues of *Danio rerio*, *Mus musculus* and *Homo sapiens* | idr0053-faas-virtualnanoscopy | Mosaic TEM | | https://doi.org/10.1083/jcb.201201140.dv |
| IDR | TEM analysis of SARS-CoV-2 infected intestine organoids | SARS-CoV-2 infected human intestine organoids | idr0083-lamers-sarscov2 | Mosaic TEM | | https://doi.org/10.17867/10000135 |
| IDR | Electron Micrographs of the nucleus | Human cultured cell | idr0086-miron-micrographs | FIB-SEM | | https://doi.org/10.17867/10000141 |
| IDR | Scanning transmission electron microscopy of Islets of Langerhans | Pancreas tissue of human type 1 diabetes | idr0116-deboer-npod | Mosaic STEM | | https://doi.org/10.17867/10000168 |
| EMPIAR | FIB-SEM of a HeLa cell | HeLa cell sample | EMPIAR-10311 | FIB-SEM (1,727 sections) | 94.0 GB (TIFF) | https://dx.doi.org/10.6019/EMPIAR-10311 |
| EMPIAR | SARS-CoV-2 infection in human adult lung alveolar stem cells | human three-dimensional alveolar type 2 cell cultures | EMPIAR-10533 | Mosaic TEM | 20.8 GB (TIFF) | https://dx.doi.org/10.6019/EMPIAR-10533 |
| EMPIAR | SARS-CoV-2 productively infects human gut enterocytes | SARS-CoV-2 infected intestine organoids | EMPIAR-10404 | Mosaic TEM | 156.2 GB (TIFF) | https://dx.doi.org/10.6019/EMPIAR-10404 |
| EMPIAR | Serial Block Face SEM of HeLa cell pellet with 10 nm pixels and 50 nm slices (benchmark dataset) | HeLa cell sample | EMPIAR-10094 | SBF-SEM (518 sections) | 129.8 GB (DM4) | https://dx.doi.org/10.6019/EMPIAR-10094 |
| EMPIAR | SBF SEM images of a Zebrafish hindbrain macrophage containing a replicating Toxoplasma gondii tachizoite | Zebrafish hindbrain sample | EMPIAR-10462 | SBF-SEM (557 sections) | 338.4 GB (DM4) | https://dx.doi.org/10.6019/EMPIAR-10462 |
| Nanotomy | Nanotomy of blistering diseases | Human autoimmune blistering disease pemphigus | | Mosaic SEM | | http://www.nanotomy.org/OA/Sokol2015JID/ |
| Nanotomy | Islets of Langerhans during Type I Diabetes | pancreas tissue in type 1 diabetic rats | | Mosaic TEM | | http://www.nanotomy.org/islets2/navigate.html |
| OpenOrganelle | Macrophage cell | Wild-type THP-1 macrophage. THP-1 human monocyte cell line (ATCC TIB-202) treated with PMA to differentiate into macrophages | jrc_macrophage-2 | FIB-SEM; Dimensions (µm): 40 x 8 x 37 (x, y, z) | | 10.25378/janelia.13123385 |
| OpenOrganelle | Killer T-Cell attacking cancer cell | OT-I mouse cytotoxic T lymphocyte attacking an ID8 cell | jrc_ctl-id8-1 | FIB-SEM; Dimensions (µm): 74 x 13 x 42 (x, y, z) | | 10.25378/janelia.13114454 |
| OpenOrganelle | Immortalized breast cancer cell (SUM159) | Wild-type SUM-159 cell, treated with 0.5 mM oleic acid for 45 mins prior to high pressure freezing to induce the formation of lipid droplets. | jrc_sum159-1 | FIB-SEM; Dimensions (µm): 64 x 11 x 35 (x, y, z) | | 10.25378/janelia.13114352 |
| BossDB | IARPA MICrONS Pinky100 | High-resolution electron microscopy, segmentation, and morphological reconstruction of cortical | pinky100 | Tape-collected section TEM; Dimensions (µm): 499.71 x 335.87 x 87.08 | | https://bossdb.org/project/microns-pinky |
| | | circuits within the visual cortex of mouse. | | | | |
| BossDB | Phelps, Hildebrand, & Graham, et al 2021 | A transmission electron microscopy dataset of the ventral nerve cord of an adult female Drosophila melanogaster. | FANC | Tape-collected section TEM; Dimensions (μm): 590.03 x 1140 x 198 | | https://bossdb.org/project/phelps_hildebrand_graham2021 |
| BossDB | Morgan et al. 2020 | Serial section electron microscopy volume of the mouse dorsal lateral geniculate nucleus (dLGN). | lgn | ATUM-SEM; Dimensions (μm): 800 x 800 x 300 | | https://bossdb.org/project/morgan2020 |
| Data information was obtained on Sep.–Nov. 2021. Abbreviation: CIL; Cell Image Library, IDR; Image Data Resource, EMPIAR; Electron Microscopy Public Image Archive, and BossDB; Brain Observatory Storage Service & Database. | | | | | | |